\documentclass[
  journal=pasa,
  manuscript=article-type,
  year=2020,
  volume=37,
]{cup-journal}

\usepackage{amsmath}
\usepackage{amssymb}
\usepackage[nopatch]{microtype}
\usepackage{booktabs}
\usepackage{orcidlink}
\usepackage{graphicx}	
\usepackage{subcaption}
\usepackage{placeins}
\usepackage{caption}
\usepackage{threeparttable}
\usepackage{hyperref}
\usepackage{bm}
\hypersetup{
    colorlinks = true,
    allcolors  = {blue} 
}
\usepackage{subcaption}
\usepackage[normalem]{ulem}
\usepackage[acronym]{glossaries}
\glsdisablehyper
\newacronym{CE}{CE}{common envelope}
\newacronym{RLOF}{RLOF}{Roche lobe overflow}
\newacronym{COM}{CoM}{centre of mass}
\newacronym{RSG}{RSG}{red supergiant}
\newacronym{RG}{RG}{red giant}
\newacronym{NS}{NS}{neutron star}

\newcommand{\msun}{M$_{\odot}$}
\newcommand{\rsun}{R$_{\odot}$}
\newcommand{\mdot}{M$_\odot$~yr$^{-1}$}
\newcommand{\phant}{{\sc phantom}}

\newcommand{\mesa}{{\sc mesa}}
\newcommand{\disk}{{\mathrm disk}}
\newcommand{\loss}{{\mathrm loss}}
\newcommand{\ini}{{\mathrm ini}}
\newcommand{\fin}{{\mathrm fin}}
\newcommand{\esc}{{\mathrm esc}}
\newcommand{\orb}{{\mathrm orb}}
\newcommand{\Lone}{{\mathrm L1}}
\newcommand{\Ltwo}{{\mathrm L2}}
\newcommand{\eff}{{\mathrm eff}}
\newcommand{\Edd}{{\mathrm Edd}}
\newcommand{\mezcal}{{\sc mezcal}}

\newcommand{\e}[1]{$\times 10^{#1}$}

\title{Three-dimensional simulations of accretion disks in pre-CE systems ---
II. Accretion efficiency and angular momentum transport
}

\author{Ana L. Juarez-Garcia}
\affiliation{School of Mathematical and Physical Sciences, Macquarie University, Balaclava Road, North Ryde, Sydney, NSW 2109, Australia}
\alsoaffiliation{Astrophysics and Space Technologies Research Centre, Macquarie University, Balaclava Road, North Ryde, Sydney, NSW 2109, Australia}
\email[Ana Juarez-Garcia]{analourdes.jurezgarca@hdr.mq.edu.au}

\author{Mike Y. M. Lau}
\affiliation{Zentrum f\"ur Astronomie der Universit\"at Heidelberg, Astronomisches Rechen-Institut, M\"onchhofstr. 12-14, 69120 Heidelberg}
\alsoaffiliation{Heidelberger Institut f\"{u}r Theoretische Studien, Schloss-Wolfsbrunnenweg 35, 69118 Heidelberg, Germany}

\author{Orsola De Marco}
\affiliation{School of Mathematical and Physical Sciences, Macquarie University, Balaclava Road, North Ryde, Sydney, NSW 2109, Australia}
\alsoaffiliation{Astrophysics and Space Technologies Research Centre, Macquarie University, Balaclava Road, North Ryde, Sydney, NSW 2109, Australia}

\author{Lionel Siess}
\affiliation{Institut d’Astronomie et d’Astrophysique, Université Libre de Bruxelles (ULB), CP 226, 1050 Brussels, Belgium}

\author{Daniel J. Price}
\affiliation{School of Physics and Astronomy, Monash University, Clayton, Vic 3800, Australia}

\author{Rebecca Nealon}
\affiliation{School of Physics and Astronomy, Monash University, Clayton, Vic 3800, Australia}

\keywords{accretion, accretion discs -- binaries (including multiple): close -- hydrodynamics -- methods: numerical -- stars: evolution} 

\begin{document}

\begin{abstract}

Rapid mass transfer preceding a \acrlong{CE} event is a critical yet poorly understood stage of binary stellar evolution, setting the initial conditions for the \acrlong{CE} inspiral and for the formation of compact binary stars. We present three-dimensional smoothed particle hydrodynamics simulations of accretion disk formation during this phase using the {\phant} code. We model the final 21~yr of \acrlong{RLOF} from a 7~\msun\ red giant onto a 1.4~\msun\ neutron star companion, with the mass transfer rate prescribed by a 1D \mesa{} model. Over this interval, the mass transfer rate increases from 1.3~\e{-4} to 1.0~\e{-1}~\mdot. An accretion disk forms around the neutron star, reading a mass of 5.0\e{-3}~\msun, a radius of 40~\rsun\ and an aspect ratio $H/R \sim 0.1$ by the end of the simulation.  A direct comparison with the grid-based simulation of Juarez-Garcia et al. (2025) shows that the two codes produce disk masses that agree to within 6\%. The accretion rate onto the neutron star reaches 5.2\e{-3}~\mdot, corresponding to 14\% of the mass injection rate and greatly exceeding the Eddington limit. We demonstrate that this accretion rate is consistent with being driven by turbulent angular momentum transport, with an effective viscosity parameter $\alpha_{\eff} = 0.03-0.06$. Ejecta leaving the binary system carries specific angular momentum approximately 90\% that of the $L_{2}$ Lagrange point, equivalent to $\sim10$ times the binary's specific orbital angular momentum. This indicates that $L_{2}$ mass loss efficiently reduces the binary orbital separation.  

 
\end{abstract}

\section{Introduction}




In a binary system the more massive star evolves first, expands, and if the companion is close enough, it starts transferring mass to the less massive star via the inner Lagrange point. Under certain conditions, mass transfer can become dynamically unstable \citep{Hjellming1987, soberman1997, Ge2010} and the binary system will undergo a \acrfull{CE} inspiral. During the \acrshort{CE} phase, the companion inspirals toward the giant star's core inside what has become a shared envelope \citep{1976IAUS...73...75P, ivanova2013common}.

High mass X-ray binaries are systems where a massive donor star \citep[M$_d$ $\geq 8$~\msun,][]{Tauris2006} transfers mass onto a neutron star or black hole companion, either by wind accretion or by Roche lobe overflow, or by wind accretion first, and Roche lobe overflow later \citep{Verbunt1993}. The accretion stream forms an accretion disk around the companion which emits X-rays. HMXBs may also go through a \acrshort{CE} interaction. 

Although it is not straightforward to determine whether an observed X-ray binary is accreting from the wind of the giant or whether the giant is overfilling its Roche lobe, it is presumed that most systems are in the former state, as this state is longer-lived. That said, there are examples of systems that are thought to actively be transferring mass via Roche-lobe overflow. An example is M33 X-7, a well-studied X-ray binary. \citet{Ramachandra22} used spectral analysis to show that the donor star is overfilling its Roche lobe. Later, \cite{Dickson2024} demonstrated, via 3D high resolution hydrodynamic simulations, that the system is transitioning from stable to unstable mass transfer. 

The evolution of the system during the pre-\acrshort{CE} envelope phase may determine whether the system will go through a CE interaction at all, or, if it does, may dictate the outcome of the interaction \citep{Iaconi2017, iaconi2018, Reichard2019}. During this phase, several physical processes occur at the same time, including the formation of an accretion disk, and possible formation of jets \citep{Armitage2000}, or the creation of outflows via the outer Lagrange points $L_{2}$ and $L_{3}$ \citep{Huang1963}. A handful of studies have modelled the formation of accretion disks and jets in pre-CE and CE systems. \citet{Shiber2019} simulated an ad hoc jet from a companion during a low mass CE interaction, showing it can help unbind envelope and affect the final binary separation. \citet{lopez2020disc} instead modelled self-regulated jets powered by accretion onto a neutron star inside a red giant's envelope and found that an accretion disk can form under these conditions. Extending this work, \citet{LopezCamara2022} simulated self-regulated jets launched by a main-sequence companion at different CE stages and found that jets do not survive the CE phase, highlighting the need for a more realistic accretion modelling.


Several other studies focused on the evolution of the binary system and its outflows. \citet{Macleod2018}, \citet{MacLeod2020} and \citet{Macleod2020b} studied the onset of mass transfer in different binary systems by evolving the donor star and assuming a point mass companion. They measured the decrease in angular momentum and orbital separation, showing that the majority of the mass and angular momentum loss from the binary occurs through the outer Lagrange points. Recently, \citet{Scherbak2025, Scherbakcooling2025} also studied the binary evolution via mass loss through the outer Lagrange points, and the effects of radiative cooling, and demonstrated that cooling does not affect the amount of angular momentum loss. All of these examples suggest that the pre-CE initial conditions directly impact the outcome of the CE, although we lack a comprehensive picture.

Recently \citet[][hereafter Paper~I]{Juarez2025}, performed hydrodynamic simulations of the formation of an accretion disk via \acrfull{RLOF} from a 7~\msun\ red giant onto a 1.4~\msun\ neutron star companion using the grid-based adaptive mesh refinement code \mezcal\ \citep{decolle2012}. They determined that a disk forms achieving a mass of $\approx 5\times 10^{-3}$~\msun, a radius of $\approx 40$~\rsun\ and a scale height of $\approx 5$~\rsun\ at the outer edge of the disk. The accretion rate onto the companion achieved a value of $\approx 4\times 10^{-3}$~\msun~yr$^{-1}$ implying an $\alpha$ parameter \citep{Shakura1973} of $\sim 0.1$ that is assumed to be driven by turbulence and shocks. Additionally, Paper~I employed the optically thin prescription of \cite{Jackson2017AStars} to define the mass injection area into the simulation, which becomes inaccurate as the donor overfills its Roche lobe.

In this paper, we carry out a similar simulation, but this time using the smoothed particle hydrodynamics (SPH) code {\phant}. In so doing, we also improve on a number of assumptions regarding the physical description of the injection stream. This allows us to carry out a side by side comparison. We also present a detailed analysis of the accretion flow within the disk, quantifying angular momentum transport through measurements of the effective viscosity parameter, $\alpha$. By considering a larger computational domain (2000~\rsun\ from the binary's initial centre of mass), we are able to study the mass loss from the binary system, quantifying the mass and angular momentum lost through the $L_{2}$ point.

This paper is organised as follows. Section~\ref{sec:method} explains the setup and gives details of the new implementation of the mass injection method. We analyse the accretion disk's evolution and parameters in Section~\ref{ssec:M_R_H}. In Section~\ref{ssec:mdot_temp_vel_alpha}, we discuss the disk's accretion rate and angular momentum transport efficiency (effective viscosity parameter). In Section~\ref{ssec:turbulence_angular_momentum}, we measure the angular momentum and mass loss rates from the binary system. We test the dependence of our results on both numerical and physical parameters in Section~\ref{sec:differences}, including code comparison with Paper~I (see Section~\ref{ssec:comp_codes}). Section~\ref{sec:conclusion} summarizes our conclusions.

\section{Methods}
\label{sec:method}

\begin{figure}[!ht]
    \centering
    \includegraphics[width=0.9\columnwidth]{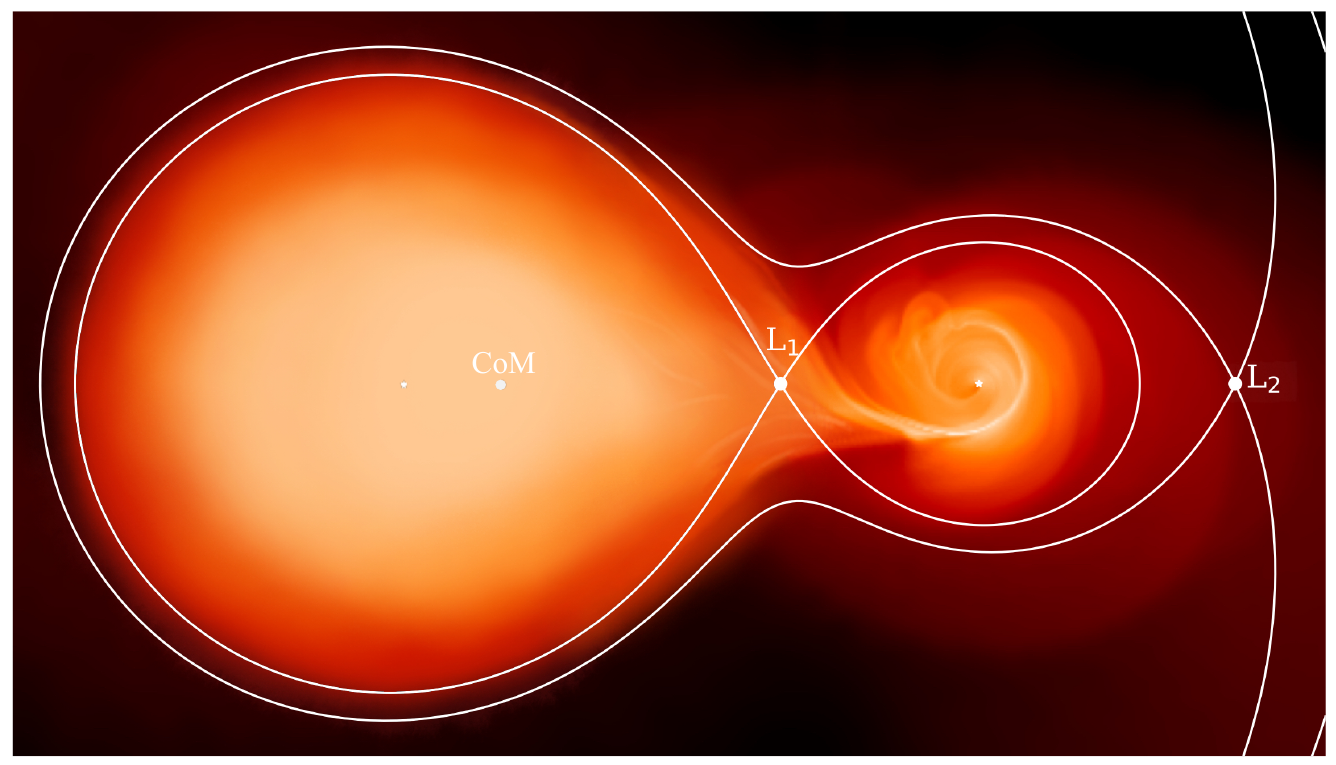}
    \caption{Artwork showing the simulation setup. The mass-transfer stream is injected at the $L_{1}$ Lagrange point towards the accretor. The donor star, a 7~\msun\ \acrlong{RG} shown on the left, is included for illustrative purposes only and is not modelled in our simulations. On the right, a 1.4~\msun\ neutron star (white star) is surrounded by an accretion disk. For reference, we plotted the isocountours of the Roche potential that include the first and second Lagrange points, $L_{1}$ and $L_{2}$. Donor star illustration by Jack Nibbs. }
    \label{fig:binary}
\end{figure}

We performed three-dimensional SPH simulations of accretion disk formation during \acrshort{RLOF}, using the SPH code \phant\ \citep{Price2018}. Our method and setup follow those of Paper~I \citep{Juarez2025}, who performed simulations of accretion disks with the finite-volume hydrodynamics code, \mezcal\ \citep{decolle2012}. The computational domain is limited by a spherical boundary with a radius of 2000~\rsun, and it is much bigger than that of Paper~I. As in Paper~I, the mass is injected through a nozzle located at the $L_{1}$ point. We consider a 1.4~\msun\ \acrfull{NS} accreting from a 7~\msun\ red giant (RG) donor. The binary separation is set to 270~\rsun\, so that the donor star just fills its Roche lobe based on \citet[][see Figure~\ref{fig:binary}]{eggleton1983}. 

\subsection{Initial parameter setup}
\label{sec:setup_1D}

The evolution of the mass transfer rate between the giant star and the compact companion was obtained by modelling the evolution of the binary system through the \acrshort{RLOF} phase using the 1D code \mesa\ (Modules for Experiments in Stellar Astrophysics; version r24.08.1; \citealt{paxton2011, paxton2013, paxton2015, paxton2018, paxton2019}; \citealt{Jermyn2023}).

The binary calculation starts with a non-rotating donor star of initial mass 7~\msun, radius 3.2~\rsun\footnote{This radius has been calculated when the hydrogen in the core has decreased by 5\%.}, and solar metallicity ($Z_{0}=0.02$), evolved from the zero-age main sequence, with a point mass \acrlong{NS} companion, with an initial binary separation of 270~\rsun\ (initial orbital period of 177 days). Convective core overshooting is not included in the donor star model, and mass loss on the red giant branch is treated using the Reimers wind prescription. Convection is described using mixing-length theory with a mixing-length parameter of $\alpha_{\rm MLT}= 1.73$. The simulation runs for $\sim 4.0 \times 10^{7}$~yr until the primary fills its Roche lobe and starts to transfer mass to the companion. At this point, the donor star is a red giant, with a helium-burning core, surrounded by a hydrogen shell undergoing CNO-cycle burning, and an extended hydrogen rich convective envelope. The orbital separation is 270.4~\rsun, the \acrlong{RG} has a mass of 6.98~\msun, a radius of 134~\rsun, an effective temperature of $3\,895$~K, a luminosity of $3\,739$~L$_\odot$, and the initial mass transfer rate is $1.42\times 10^{-8}$~\mdot.

\begin{figure}[ht!]
    \centering
    \includegraphics[width=0.85\columnwidth]{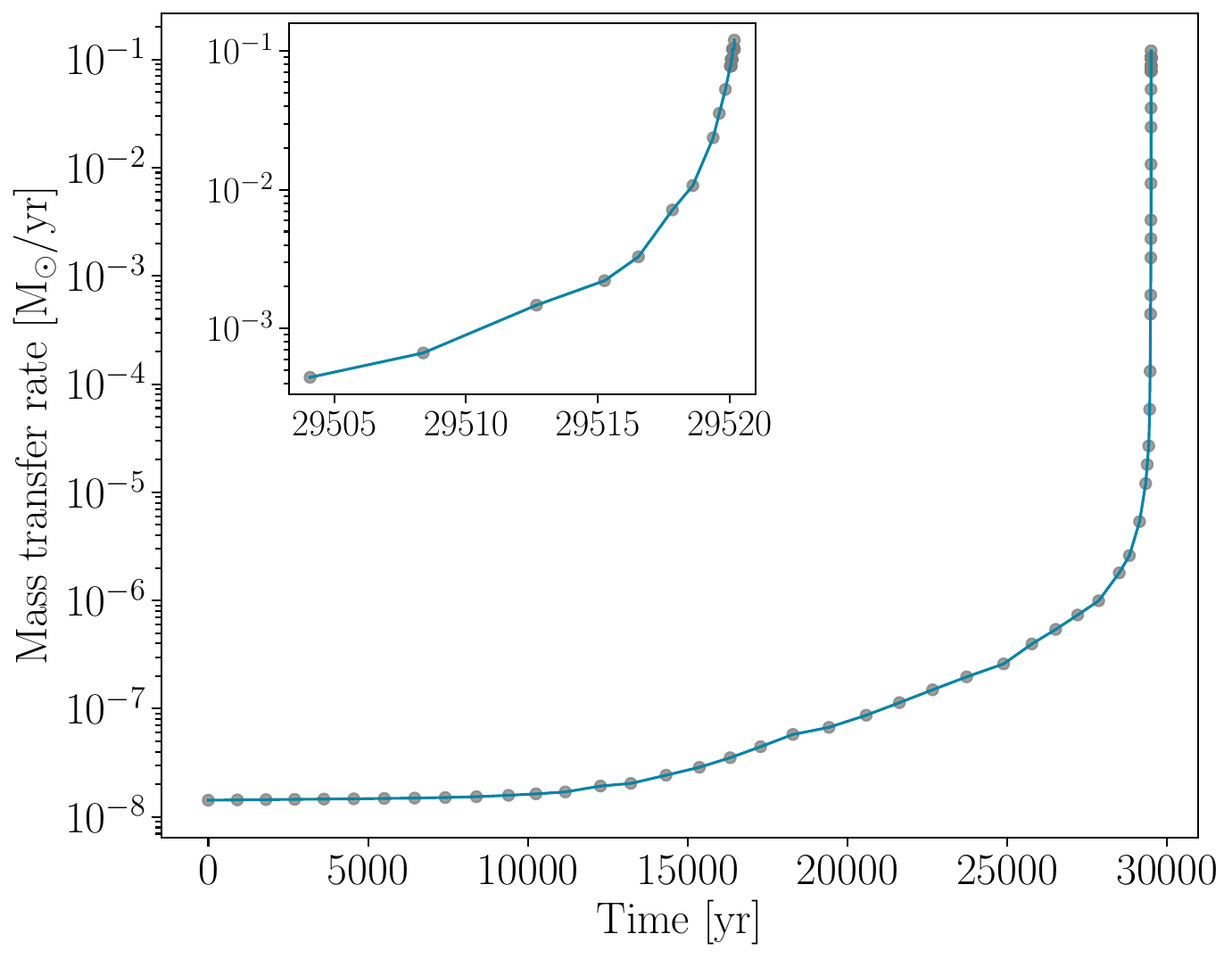}
    \caption{Mass transfer computed by the \mesa\ simulation code. The inset shows the last 21 yr of mass transfer, which we simulate in 3D using \phant.}
    \label{fig:mdot_mesa}
\end{figure}

We compute the mass transfer rate using the \cite{Kolb90} scheme, which accounts for an optically thick stream when the Roche radius is located below the donor's photosphere. The mass transfer parameters $\alpha_{\rm mt}$, $\beta_{\rm mt}$, and $\delta_{\rm mt}$ (equation 5 of \citealt{paxton2015}) are left at their default value of zero, indicating that mass transfer is fully conservative. The simulation continues for an additional $\approx 30\,000$~yr, during which the mass transfer rate increases to 1.03$\times10^{-1}$~\mdot\ (Figure~\ref{fig:mdot_mesa}). After 30\,000~yr, the red giant has a mass of 6.89~\msun, a radius of 144~\rsun\ and an effective temperature of $\sim5\,110$~K, while the companion star's mass has increased to 1.49~\msun\ through accretion. Over the mass transfer phase, the orbital separation has decreased to 242~\rsun. Beyond this point, the mass transfer rate increases rapidly enough that we assume it quickly leads to a \acrshort{CE}. 

It is computationally unfeasible to simulate the whole mass transfer phase in 3D, so following Paper~I, the 3D SPH simulation only simulates the last 21~yr (see inset in Figure~\ref{fig:mdot_mesa}). The time simulated in 3D is from 29\,500~yr to 29\,521~yr in the 1D \mesa\ simulation, during which the conservative mass transfer rate increases from 1.31\e{-4}~\mdot\ to 1.03\e{-1}~\mdot. By the start of our 3D simulation, the donor star has a mass of 6.97~\msun\ and a radius of 139.15~\rsun. At this stage, the companion star has a mass of 1.41~\msun, the orbital separation has decreased slightly to 266.91~\rsun, and the orbital period is 174.4 days. The 1D MESA model parameters at this moment are used to set up our 3D simulation. 


\subsection{3D numerical setup}
\label{sec:setup_3D}


The 3D simulation consists of two point-mass gravitational potentials fixed in the corotating frame with masses $M_{d}~=~6.97$~\msun\ and $M_{a}~=~1.41$~\msun, representing the donor and companion (accretor) star respectively, with an orbital separation of $a~=~266.91$~\rsun. SPH particles are deleted when they enter the accretion radius $r_{a}~=~1$~\rsun\ around the companion, to simulate accretion. A very small fraction of the material injected at $L_{1}$ flows back into the donor's Roche lobe, so we also remove particles that are within $r_{d}~=~100$~\rsun\ of the centre of the donor. The simulation domain extends up to a radius of 2000~\rsun\ from the system's \acrfull{COM}, beyond which particles are removed from the simulation. 

We inject mass through a ``nozzle'' located at the position of $L_{1}$, $x_{\Lone}~=~130.81$~\rsun\ relative to the frame's origin at the \acrshort{COM} of the system. The injection nozzle is represented by a close-packed lattice of particles, with its axis aligned along the line connecting the two stars ($x$-axis). Particles are injected into concentric circular layers whose radii evolve over time. Within each layer, particles maintain fixed relative positions and are assigned a prescribed injection velocity. In \ref{app:appendix_radius} we give details of the injection setup. The setup configuration of {\phant} is denoted \verb+masstransfer+ and is available in the public code.

\begin{figure}[ht!]
    \centering
    \includegraphics[width=\columnwidth]{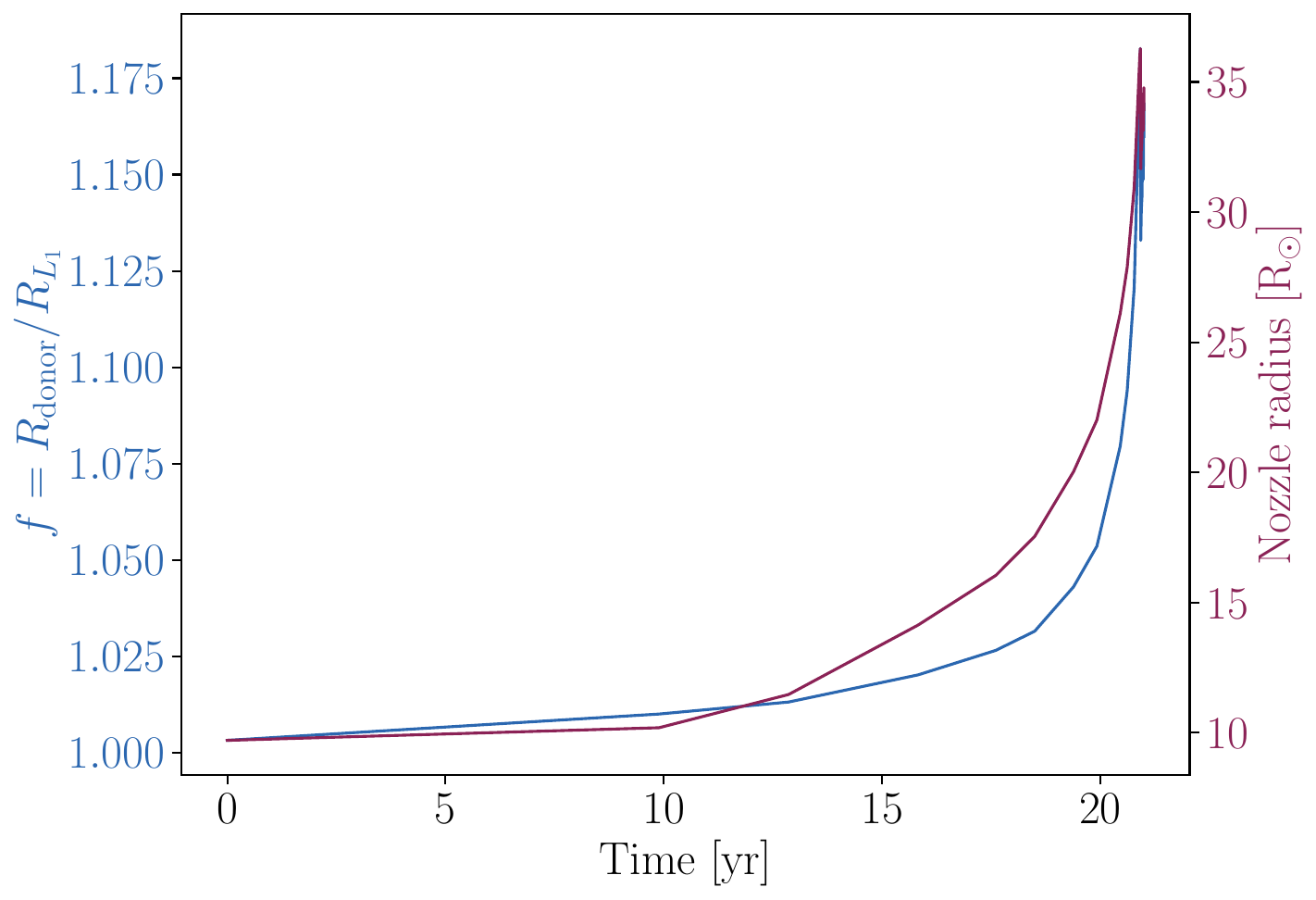}
    \caption{Evolution of the donor's Roche lobe overfilling factor (blue line) and nozzle radius (magenta line) as calculated by the 1D \mesa\ simulation and used in the 3D \phant\ simulations.}
    \label{fig:radius}
\end{figure}

In the 1D simulations, the mass transfer rate is calculated using the prescription of \citet{Kolb90}, which accounts for the transition from an initially optically thin flow, described by \citet{Ritter1988}, to a later optically thick regime. During the simulated mass-transfer phase, the donor overfills its Roche lobe by up to $\sim 18$\% (Figure~\ref{fig:radius}), justifying the use of the optically thick mass transfer prescription of \citet{Kolb90}. In this formalism, the radius of the nozzle grows from 9.71~\rsun\ to 33.9~\rsun, an effect that was not considered in Paper~I, where the optically thin prescription of \cite{Jackson2017AStars} was used to estimate the mass transfer rate. So, in our simulations, the nozzle radius and the mass transfer rate are imposed as input parameters in the three‑dimensional setup and their evolution is presented in Figure~\ref{fig:radius}. The number density and velocity of particles are set to obtain the mass transfer rate as prescribed by the 1D \mesa\ simulation. Additional details of the numerical setup of the nozzle are provided in \ref{app:appendix_radius}.

The density in the nozzle is given by $\rho_{\Lone} = \dot{M}_{\Lone}/(S_{\Lone}v_{\Lone})$, where $\dot{M}_{\Lone}$ is the mass transfer rate from the 1D simulation, $v_{\Lone}$ the velocity of the injected stream, and $S_{\Lone}$ the area of the injection region given by \mesa. 
As the donor star expands and overflows its Roche lobe, the mass transfer stream becomes supersonic after passing through $L_1$, with the sonic point expected to lie near $L_1$ \citep{lubow1975gas}. Motivated by this, and as an improvement over Paper~I where the gas was injected with subsonic velocity ($v_{\Lone} = 0.1 c_{\rm s}$), we here set the velocity of injection to be the sound speed along the $x$-axis: $v_{\Lone}~=~(c_{\rm s},0,0)$, where the sound speed is $c_{\rm s}$~=~6.42~km~s$^{-1}$ considering an average gas temperature of 3\,000~K given by the donor's effective temperature during the 21 yr of evolution. The pressure of the injected mass is given by $P_{\Lone} = c_s^2 \rho_{\Lone}/\gamma$, where $\gamma=1.1$. 

As we inject mass into the domain, the number of SPH particles continuously grows and eventually becomes computationally expensive. To address this, we apply the particle merging scheme of \citet{Nealon25} at selected times, constraining the total number of particles to remain below 2 million. We perform a global merge of particles every time the total number of particles reaches 2 million, resulting in 1 million particles immediately after. During each merging event, the individual particle mass, $m_{\rm part}$, increases by a factor of two. For further details, see \ref{app:merging}, where we carry out simulations with three initial particle masses ($10^{-10}$, $10^{-11}$ and $10^{-12}$~\msun), and perform a number of checks to ensure that particle merging has no impact on the results. 

We assume an ideal gas equation of state
\begin{equation}
    P = \left(\gamma - 1\right)\rho u,
    \label{eq:idealgas}
\end{equation}
where the gas pressure $(P)$ is determined by the gas density $(\rho)$ and internal energy $(u)$, with a fixed adiabatic index of $\gamma=1.1$, which is close to the isothermal limit in order to represent efficient cooling \citep{makita2000two, Huarte2013, Murguia-Berthier2017AccretionFormation, macleod2017common, Macleod2018, MacLeod2020}. 
We neglect self-gravity in the simulations because the total mass of the disk is much smaller than that of the binary system. Since the gravitational force from each star scales with its mass, the mutual gravitational force between gas particles in the disk is negligible (up to 12 orders of magnitude smaller) compared to the gravitational force exerted by the point mass potentials.

\section{Results}
\label{sec:results}
\begin{figure*}
\centering
  \includegraphics[width=\linewidth]{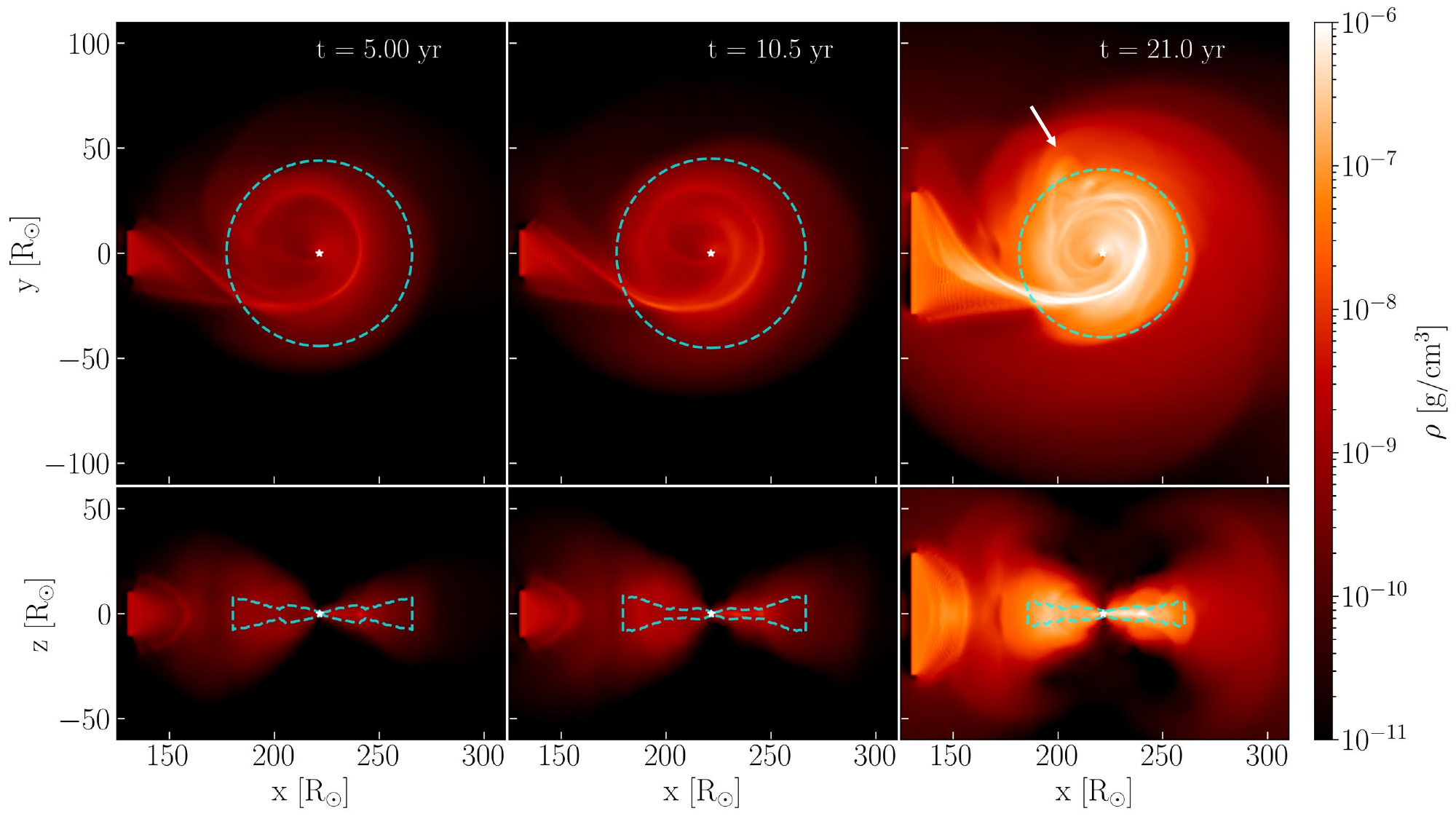}
\caption{Density cross-sections of the fiducial simulation in the orbital plane (upper panels) and edge-on plane (bottom panels) at 3 different times in the simulation. The slices are aligned with the accretor star's position ($x_{a}$ = 221.53~\rsun), indicated by the central white star. The blue dashed lines indicate the disk's radius and scale height, the white arrow in the top right panel shows a local over-density (see text). Animations showing the full temporal evolution of these density slices are available at Zenodo: doi:\url{10.5281/zenodo.20710411}.}
\label{fig:cross_sections_density}
\end{figure*}

In this section, we present the results of our fiducial simulation ({\it sim-fiducial} in Table~\ref{tab:simulations}) carried out with \phant. As we describe our results, we will compare this simulation with that of Paper~I, and isolate the impact of the simulation updates in Section~\ref{sec:differences}. A comprehensive summary of the input parameters for all simulations is provided in  Table~\ref{tab:simulations}. 

\subsection{Disk properties}
\label{ssec:M_R_H}
\begin{table*}
    \centering
    \begin{tabular}{lcccccc}
    \hline
    \hline
         Simulation &  Nozzle radius&  $\mathcal{M}_{\Lone}$&  $\dot{M_{\ini}}$ &$\dot{M}_{\fin}$& Simulated time &Adiabatic index\\
 & [\rsun]& & [\mdot] &[\mdot] &[yr]&$\gamma$\\
    \hline
        sim-fiducial& 9.71 - 33.9& 1&  1.31\e{-4}&1.03\e{-1}&21&1.1\\
         sim-compare&  12.2&  0.1&   2.30\e{-4}&9.70\e{-2}& 21&1.1\\
         sim-mach-01&  9.71 - 33.9&  0.1&   1.31\e{-4}&1.03\e{-1}& 21&1.1\\
        sim-rad-nozzle& 12.2& 1& 1.31\e{-4}& 1.03\e{-1}& 21&1.1\\
        sim-10yr& 14.1 - 33.9& 1& 1.47\e{-3}& 1.03\e{-1}& 10&1.1\\
        sim-45yr& 9.25-33.9& 1& 5.84\e{-5}& 1.03\e{-1}& 45&1.1\\
        sim-gam101& 12.2& 1& 1.31\e{-4}& 1.03\e{-1}&21&1.01\\
        sim-gam133& 9.71 - 33.9& 1& 1.31\e{-4}& 1.03\e{-1}&21&1.33\\
        sim-gam167& 9.71 - 33.9& 1& 1.31\e{-4}& 1.03\e{-1}&21&1.67\\    
 \hline
    \end{tabular}
    \caption{Input parameters for all simulations presented in this work. Column (2) lists the injection nozzle radius range, while column (3), $\mathcal{M}_{\rm  L1}$, gives the Mach number of the injected flow,  $\mathcal{M}_{L1} = v_{\Lone}/c_{\rm s}$. The initial and final mass transfer rates, $\dot{M}_{\ini}$ and $\dot{M}_{\fin}$, are given in columns (4) and (5), respectively. Column (6) lists the total simulated time, and column (7) gives the adiabatic index ($\gamma$). Other than {\it sim-10yr} and {\it sim-45yr}, all simulations initially have a 6.98~\msun\ donor star and a 1.41~\msun\ accretor, covering the final 21~yr of the mass transfer phase. The sound speed of the injected mass is set to c$_{\rm s}$~=~6.43~km~s$^{-1}$, with an initial particle mass of $m_{\rm part} = 1\times10^{-10}$~\msun.}
    \label{tab:simulations}
\end{table*}

In Figure~\ref{fig:cross_sections_density}, we show the density in the orbital and meridional planes at three times during the disk's evolution. The slices are aligned with the accretor position ($x_a$~=~221.53~\rsun); blue dashed lines indicate the radius and scale height of the disk, calculated using the methodology described below. As gas is injected into the domain, it leaves the nozzle and freely falls towards the accretor with supersonic velocities \citep{lubow1975gas}. The injection stream is initially ballistic, with its trajectory subsequently bending under the influence of Coriolis and centrifugal forces within the Roche potential \citep{frank1992book}. Eventually the stream self-intersects, producing shocks that convert kinetic energy into thermal energy; it then settles into orbits around the circularisation radius ($R_{\rm circ} =$ 21~\rsun). Subsequent shocks in the flow (visible as spiral structure) redistribute angular momentum, spreading the material and ultimately leading to the formation of an accretion disk (see top panels of Figure~\ref{fig:cross_sections_density}).  

The incoming stream interacts with the disk material creating a bow shock, as seen in the edge-on view (bottom panels of Figure~\ref{fig:cross_sections_density}, at $x\sim150$~\rsun). Throughout the evolution, the injected material maintains a distinct high-density spiral structure on the orbital plane. Over time, the disk grows and spreads radially and vertically. We also observe denser structures changing position and shape over time (Figure~\ref{fig:cross_sections_density}, top right panel, white arrow), developing at the interface between the disk and the surrounding, less dense gas, suggesting the presence of shear features. 

To measure the radius, $R_{\disk}$, mass, $M_{\disk}$, and scale height, $H(R_{\disk})$ of the accretion disk, we follow a similar method to that used in Paper~I in order to allow a direct comparison, with improvements by \cite{Lee2022} and \cite{Malfait24}. Unless otherwise specified, all accretion disk properties are measured in a coordinate system centred on the accretor star. For an isothermal disk in hydrostatic equilibrium, the scale height represents the vertical distance from the mid-plane where the density drops by a factor of $1/\sqrt{e}$, following the profile
\begin{equation}
    \rho(z) = \rho_0 e^{-\frac{1}{2}(z/H)^2}.
    \label{eq:rho_hyrdro}
\end{equation}
We divide the computational domain around the accretor star into 100 radial bins out to 90~\rsun\ (just before reaching the nozzle) and 10 azimuthal sectors ($\phi$). For each radial-angular segment ($r_{i},\phi_j$), we identify the maximum density value, $\rho_0(r_{i},\phi_j)$, usually located in the mid-plane. Since the disk is not symmetric across the orbital plane, we determine the scale height independently for both positive and negative $z$-directions by finding the point where $\rho(z = H) = \rho_0(r_i,\phi_j)/\sqrt{e}$. These two values are averaged to obtain a single scale height, $H(r_{i},\phi_j)$, for the segment. Finally, we take the median of the ten azimuthal segments at each radius to derive the azimuthally averaged scale height profile, $H(r_{i})$.

To determine the mass of the disk, we calculate the cumulative mass profile, $M(r_i)$, for half of the disk ($\phi \in [-\pi/2, \pi/2]$), to exclude the influence of the injection stream. This calculation is performed within cylindrical shells restricted to $|z|< H(r_i)$. We define the disk's radius, $R_{\disk}$, as the inflection point where the cumulative mass curve flattens out. The total disk mass, $M_{\disk}$, is then defined as twice the cumulative mass measured at  $R_{\disk}$, within the scale height. 

In the first column of Table~\ref{tab:disk_parameters}, we list the disk's properties at 21~yr. For comparison, the main results of Paper~I are listed in the second row. Although the total mass and radial extent of the disk are consistent with Paper~I, our fiducial simulation produces a disk that is more vertically compressed with a significantly higher temperature at the edge of the disk. The measured disk's radius of $\sim$40~\rsun\ is smaller than the expected truncation radius of $R_{\rm t}\approx 60$~\rsun\ \citep{paczynski1977model}, this discrepancy can be attributed to our method for determining the disk's radius in this paper and Paper~I. 
In Section~\ref{ssec:comp_codes} and \ref{ssec:comp_nozz} we will analyse whether these differences are due to the numerical code (third column of Table~\ref{tab:disk_parameters}) or to the improvements in the setup. 

\begin{figure}
    \centering
    \includegraphics[width=0.85\columnwidth]{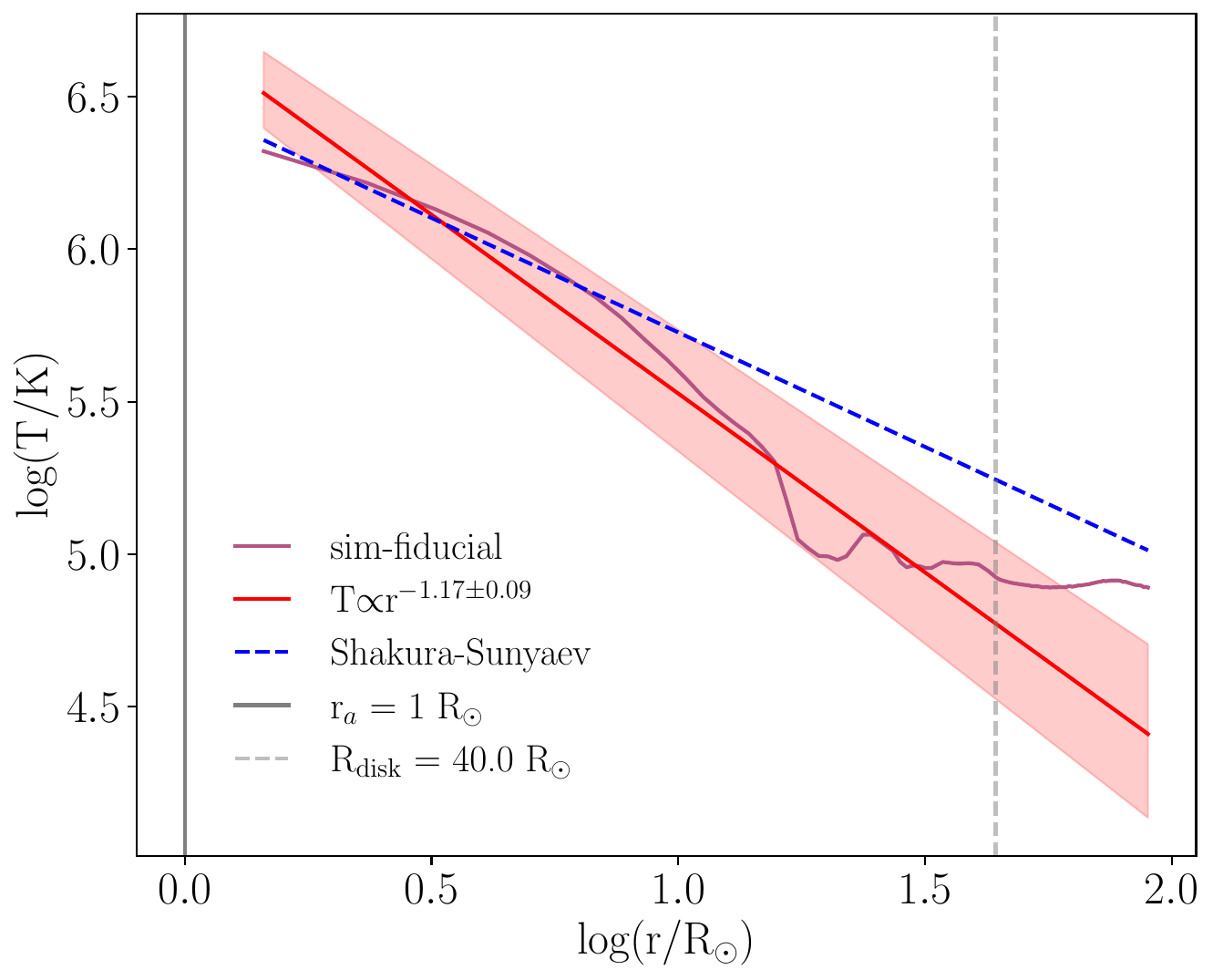}
    \caption{Temperature profile at 21 yr (magenta line). The solution for a steady thin disk \citep{Shakura1973} is shown by the blue line (T~$\propto r^{-3/4}$) and the best fit, given by $\alpha = 1.17 \pm 0.09$, is shown by the red line and red shaded area. As reference, we indicate the disk's radius with a vertical grey dashed line, and the accretion radius of the companion star with a vertical solid line.}
    \label{fig:temperature_results}
\end{figure}    

In Figure~\ref{fig:temperature_results} we present the disk's temperature profile. To measure it, we consider only half of the disk ($\phi \in [-\pi/2, \pi/2]$) thus avoiding the injection stream. Near the accretor, the mean temperature is $\sim$2\e{6} K, similar to the temperature reported in Paper~I ($\sim$10$^{6}$~K) at the same location. At the outer edge of the disk ($R_{\disk} = 40$~\rsun), the mean temperature is 93\,000~K, higher than the 60\,000~K reported in Paper~I. A more detailed discussion of these results is provided in Section~\ref{sec:differences}. The temperature profile follows a power-law relation, $T~\propto r^{-1.16\pm0.06}$. The uncertainty (red line and shaded area in Figure~\ref{fig:temperature_results}) represents the 95\% confidence interval of the linear regression in log-log space. For reference, we include the temperature profile solution for a standard thin disk model \citep{Shakura1973}, $T\propto r^{-3/4}$, effectively assessing to what extent our adopted $\gamma=1.1$ approximates the properties of a radiating disk.

 \begin{figure}
    \centering
    \includegraphics[width=0.85\columnwidth]{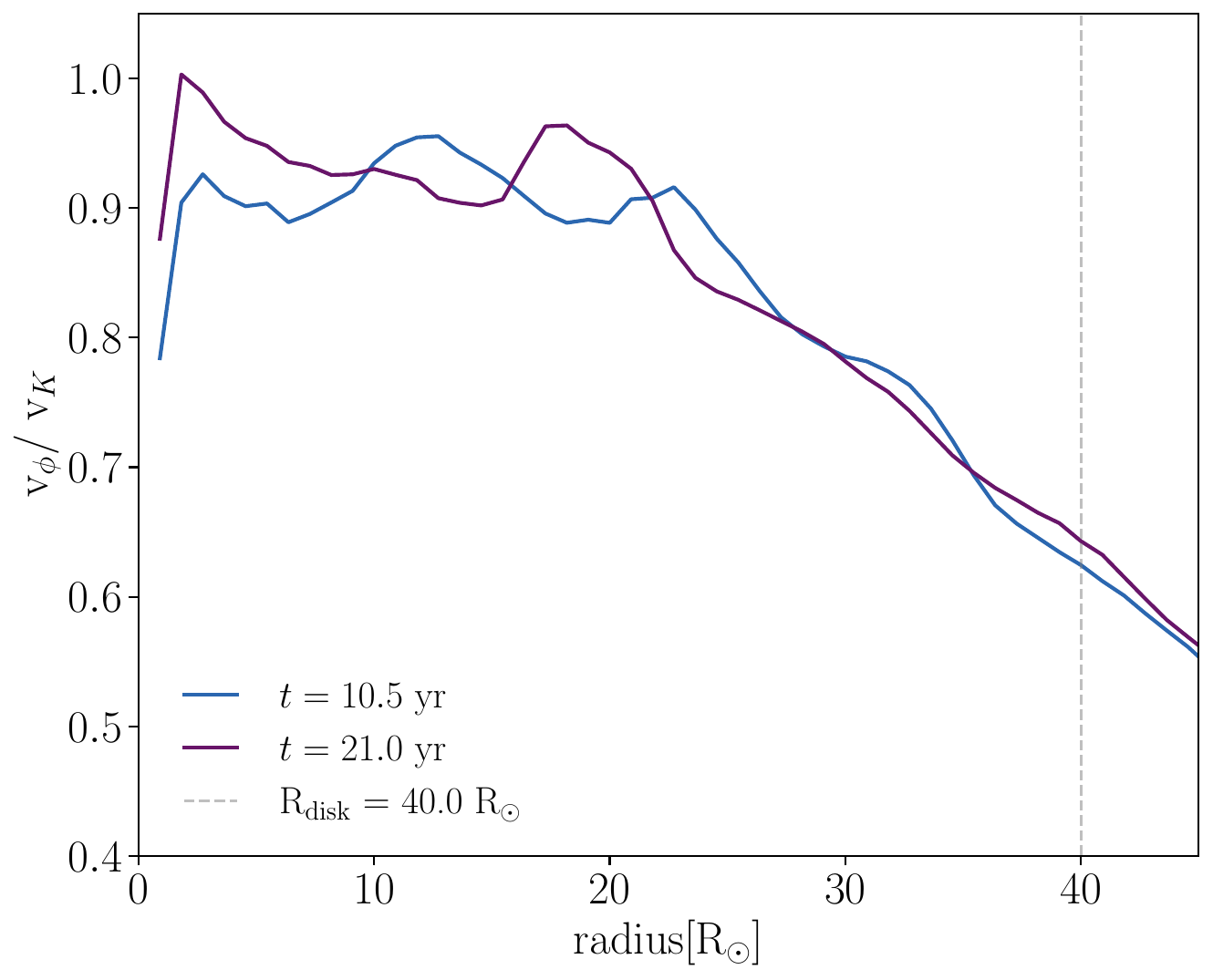}
    \caption{Azimuthal velocity profile normalised by the Keplerian velocity $(v_{\rm K})$ for two different times during the evolution of the accretion disk.}
    \label{fig:vphi_simulation}
\end{figure}    

Finally, in Figure~\ref{fig:vphi_simulation}, we plot the azimuthal velocity ($v_\phi$) of the disk normalised by the Keplerian value ($v_{\rm K} = \sqrt{GM_a/r}$) at 10.5 yr (blue line) and at 21 yr (purple line). At 21~yr, the velocity of the particles decreases with radii from $v_{\phi} = 486.9$~km s$^{-1}$ at $r=1$\rsun\ to $v_{\phi} = 53.9$~km s$^{-1}$ at  $r=40$~\rsun. At 10.5~yr and 21~yr, the disk is sub-Keplerian throughout, remaining close to unity ($\sim 0.89~v_{\rm K}$ at 10.5~yr and $\sim 0.93~v_{\rm K}$ at 21~yr) up to the circularisation radius (21~\rsun), and decreasing outwards beyond this radius. At the outer edge of the disk ($R_{\disk}=40$~\rsun), the azimuthal velocity is $\sim 0.64~v_{\rm K}$ in both profiles, indicating moderate pressure support at large radii. These values are consistent with figure~11 in Paper~I.

\subsection{Accretion rate and angular momentum transport}
\label{ssec:mdot_temp_vel_alpha}

\begin{table*}[ht!]
    \centering
    \begin{tabular}{lccccc}
    \hline
    \hline
          &  SPH simulation& Paper~I results$^a$ &SPH simulation same as Paper I &$\mathcal{M}_{\Lone}=0.1$ & optically thin nozzle simulation\\
 & (sim-fiducial)& & (sim-compare) &(sim-mach-01) &(sim-rad-nozzle)\\
    \hline     
Disk's mass [\msun]& 5.0\e{-3}&5.5\e{-3}&5.4\e{-3}&4.3\e{-3} &6.1\e{-3}\\
 Disk's radius [\rsun]& 40& 39&40&45 &42\\
 Disk's scale height [\rsun] & 5.3& 4.9&5.0&8.6 &5.2\\
 Temperature at $R_{\disk}$ [K] & 93\,000& 60\,000&71\,000 &95\,000 &92\,000\\
 Mass accretion rate [\mdot]& 5.2\e{-3} & 3.7\e{-3} & 6.0\e{-3}  &6.0\e{-3} &4.4\e{-3}\\
 \hline
 \multicolumn{4}{l}{$^a$We are referring to sim-0 in that paper.} &
  &\end{tabular}

    \caption{Disk properties at 21~yr for the fiducial simulation (sim-fiducial), for the fiducial simulation in Paper~I (sim-0 in that paper), and for an exact replica of sim-0, but carried out with \phant\ (sim-compare), for the simulation in the optically thick regime with $\mathcal{M}_{\Lone}=0.1$ (sim-mach-01), and simulation following the optically thin regime with $\mathcal{M}_{\Lone}=1.0$ (sim-rad-nozzle).}
    \label{tab:disk_parameters}
\end{table*} 

\begin{figure}[ht!]
    \centering
    \includegraphics[width=0.85\columnwidth]{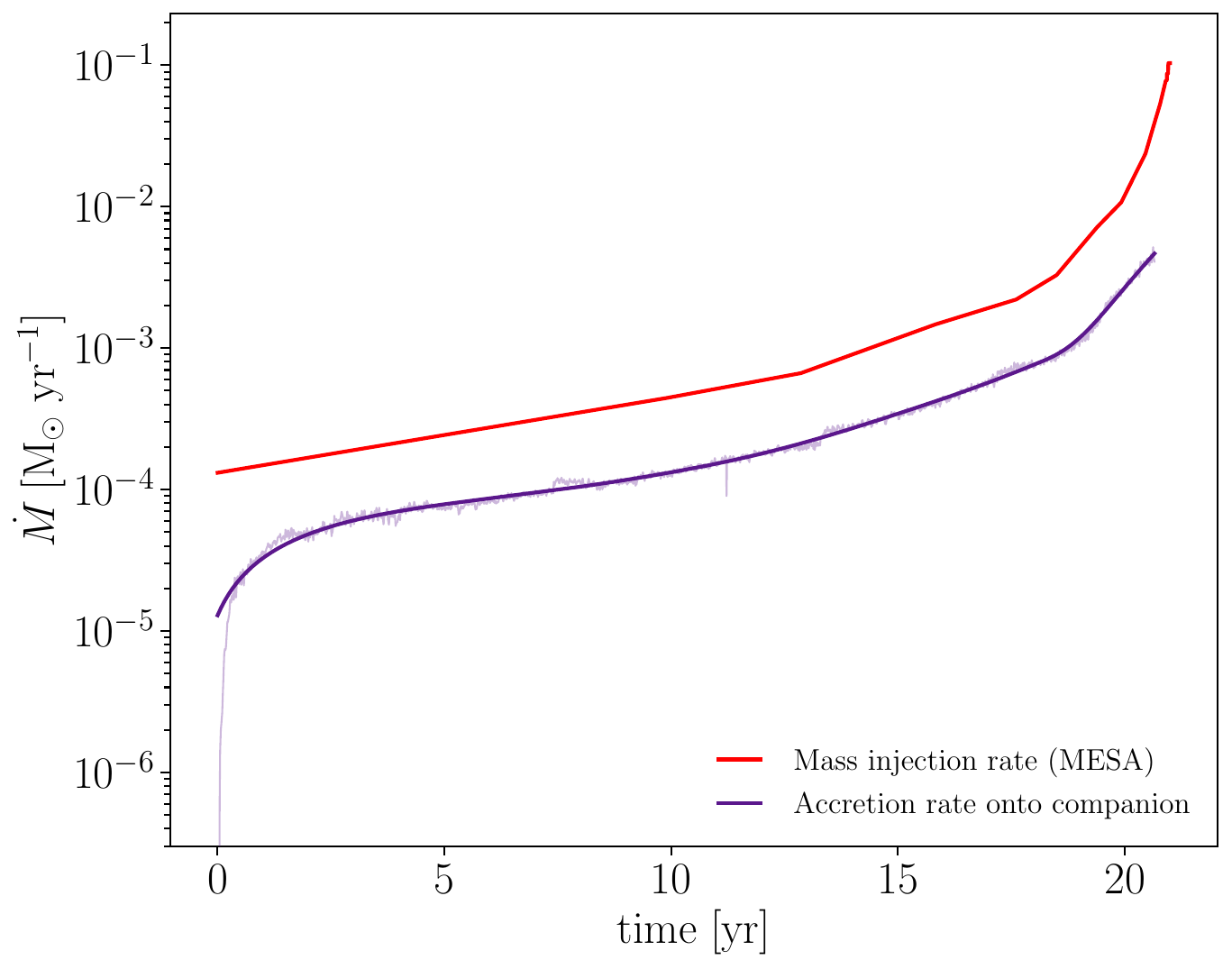}
    \caption{Mass transfer rate versus time. Red line: mass transfer rate prescribed by the \mesa\ simulations. Purple line: mass accretion rate onto the companion star.  The difference between the total injected mass and the accreted mass is attributed to mass lost in the outflow and stored in the accretion disk.}
    \label{fig:mdot_simulation}
\end{figure}

Figure~\ref{fig:mdot_simulation} shows the accretion rate onto the neutron star companion (purple line) reaching a value of 5.2\e{-3}~\mdot\ by 21~yr, which is much greater than the Eddington limit of $\dot{M}_{\Edd}\sim 2\times10^{-8}$~\mdot\ for a 1.41~\msun\ neutron star \citep{lopez2020disc}. The total accreted mass is 8.4\e{-3}~\msun, corresponding to 14\% of the total injected mass (0.061~\msun, red line in Figure~\ref{fig:mdot_simulation}). The remaining injected gas is either stored within the accretion disk or lost via outflows. We present a detailed analysis of the mass-loss rate in the binary outflow in Section~\ref{ssec:turbulence_angular_momentum}. 

Such strongly super-Eddington accretion would likely produce X-ray feedback and launch relativistic jets. Assuming that 10\% of the accreted mass is redirected into a pair of jets \citep{Fender2010, Nisini2018}, which corresponds to a jet mass-loss rate of $\dot{M}_{\rm jet} =$~5.2\e{-4}~\mdot, we calculate two possible mechanical luminosities. If the jet is launched at the accretion radius, $h_{a} = 1$~\rsun\ (with an escape velocity of $v_{\esc} = 733$~km s$^{-1}$), the mechanical luminosity is 2.3\e{4}~L$_\odot$. However, if we assume the jet is launched from the neutron star surface ($v_{\esc}\sim~0.3~c$), the mechanical luminosity is 3.5\e{8}~L$_\odot$. The latter value is consistent with the $10^{6}$ to $10^{8}$~L$_\odot$ range observed in ultraluminous X-ray sources \citep{Bachetti2014}.

\begin{figure*}[ht!]
    \centering
    \includegraphics[width=\columnwidth]{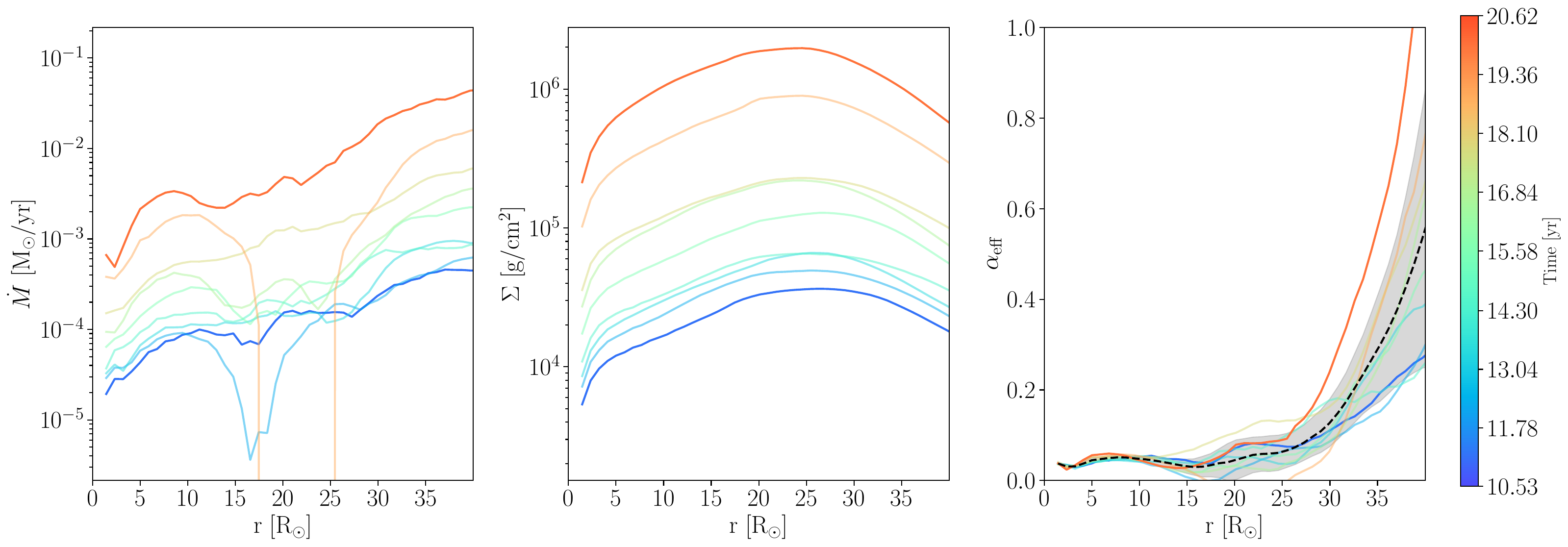}
    \caption{Temporal evolution of selected disk properties during the last 10 years of the simulation. Left panel: mass-transport rate, $\dot{M} (r)$, where negative values indicate outward flux. Middle panel: the surface density profile, $\Sigma (r)$. Right panel: effective viscosity parameter, $\alpha_{\eff} (r)$ (Eq.~\ref{eq:alpha_ss}). The colour scale indicates the time of the profile; the dashed black line and gray shaded area in the right panel represent the time-averaged $\alpha_{\eff}$ and its 1$\sigma$ standard deviation. Each colour curve is itself a short-term average of 10 individual profiles (model outputs), taken over a short time span of 0.5 years. We have plotted the profiles up to the disk's radius, $R_{\disk}$=40~\rsun. The radial distance is measured from the NS.}
    \label{fig:alpha}
\end{figure*}    

To understand the mechanism responsible for mass accretion, we compute the effective viscosity parameter, $\alpha_\mathrm{eff}$, based on the radial mass-transport rate through 100 concentric cylindrical shells within the local scale height, $H(r)$. If we assume the mass-transport rate between shells can be approximated by the prescription for geometrically thin, steady disks by \citet{Pringle1981}, and \citet{frank1992book}, $\alpha_{\eff}(r)$ is defined as
\begin{equation}
    \alpha_{\eff}(r) = \frac{\dot{M}(r)}{3\pi\Sigma(r) c_{\rm s}(r)H(r)} \left(1-\sqrt{\frac{r_{a}}{r}}\right),
    \label{eq:alpha_ss}
\end{equation}
where $\dot{M}(r)$ is the mass-transport rate, $r_{a}~=~1$~\rsun\ is the companion star's accretion radius, $\Sigma(r)$ is the surface density, and $c_{\rm s}(r)$ is the sound speed. In Figure~\ref{fig:alpha}, we present the temporal evolution of the radial profiles of the mass-transport rate ($\dot{M}$, left panel), surface density ($\Sigma$, middle panel), and the resulting effective viscosity parameter ($\alpha_{\eff}$, right panel). Short-term averages over $\delta t~=~0.5$~yr are used because the disk properties exhibit large stochastic fluctuations on shorter timescales. 
Two important times during the evolution of the disk are highlighted: the solid dark blue line at $t~=~10.5$~yr represents a time during which the mass transfer rate ($\dot{M}_{\Lone}$) remains approximately constant over the $\delta t =0.5$~yr, while the solid orange line at $t~=~21$~yr captures the disk during a phase of exponential increase in the gas injection rate. In the right panel, the dashed black line represents the time-averaged $\alpha_{\eff}$ calculated over the final ten years of the simulation.

In Figure~\ref{fig:alpha}, the mass-transport rate (left panel) increases radially outward, with negative values observed between 15~-~25~\rsun\ at $t=19.36$~yr, indicating localised regions where material is moving outward. The surface density profile (middle panel) peaks within a radial range of $20-30$~\rsun, corresponding to where the stream penetrates into the disk, also coincidental with the circularisation radius. As the mass transfer rate increases with time, both $\dot{M}$ and $\Sigma$ grow at all radii, but the overall shape of the profiles is preserved. Although both quantities increase with time, the radial profile of the effective viscosity parameter $\alpha_{\eff}$ (right panel) remains approximately flat for $r\lesssim 20$~\rsun, because the simultaneous increase in $\dot{M}$ and $\Sigma$ (Eq.~\ref{eq:alpha_ss}) largely compensates for each other. 
At $t=20.62$~yr, the mass transfer rate increases exponentially, causing $\alpha_{\eff}$ (orange line) to increase at radii larger than $r\sim25$~\rsun. Below this radius, no clear specific trend is observed in the $\alpha_{\eff}$ profile. We ascribe this effect to the impact stream penetrating the disk at $r\sim 25$~\rsun\, directly advecting angular momentum inwards rather than turbulent transport.
We therefore restrict our analysis to $r\leq25$~\rsun, where the disk structure is not directly influenced by the stream impact.

Evidence of the interaction between the disk and the incoming gas stream can be seen in the density cross-sections along the orbital plane (Figure ~\ref{fig:cross_sections_density}). It is clear from the figure that the stream penetrates down to a relatively small radius, a similar interaction was observed in the RLOF simulations of \citet{Fujiwara2001}. The disk's resultant hydrodynamic turbulence explains the fluctuations in the $\dot{M} (r)$, $\Sigma$, and $\alpha_{\eff}$ profiles shown in Figure~\ref{fig:alpha}, including the occasional negative values of the mass-transport rate. For smaller disk radii, 
despite some variability, the azimuthally averaged radial velocity, $\langle v_r\rangle$, is directed inward, consistent with the relatively constant values of $\alpha_{\eff}$ for $r \leq 25$~\rsun.  

\begin{figure}
    \centering
    \includegraphics[width=\columnwidth]{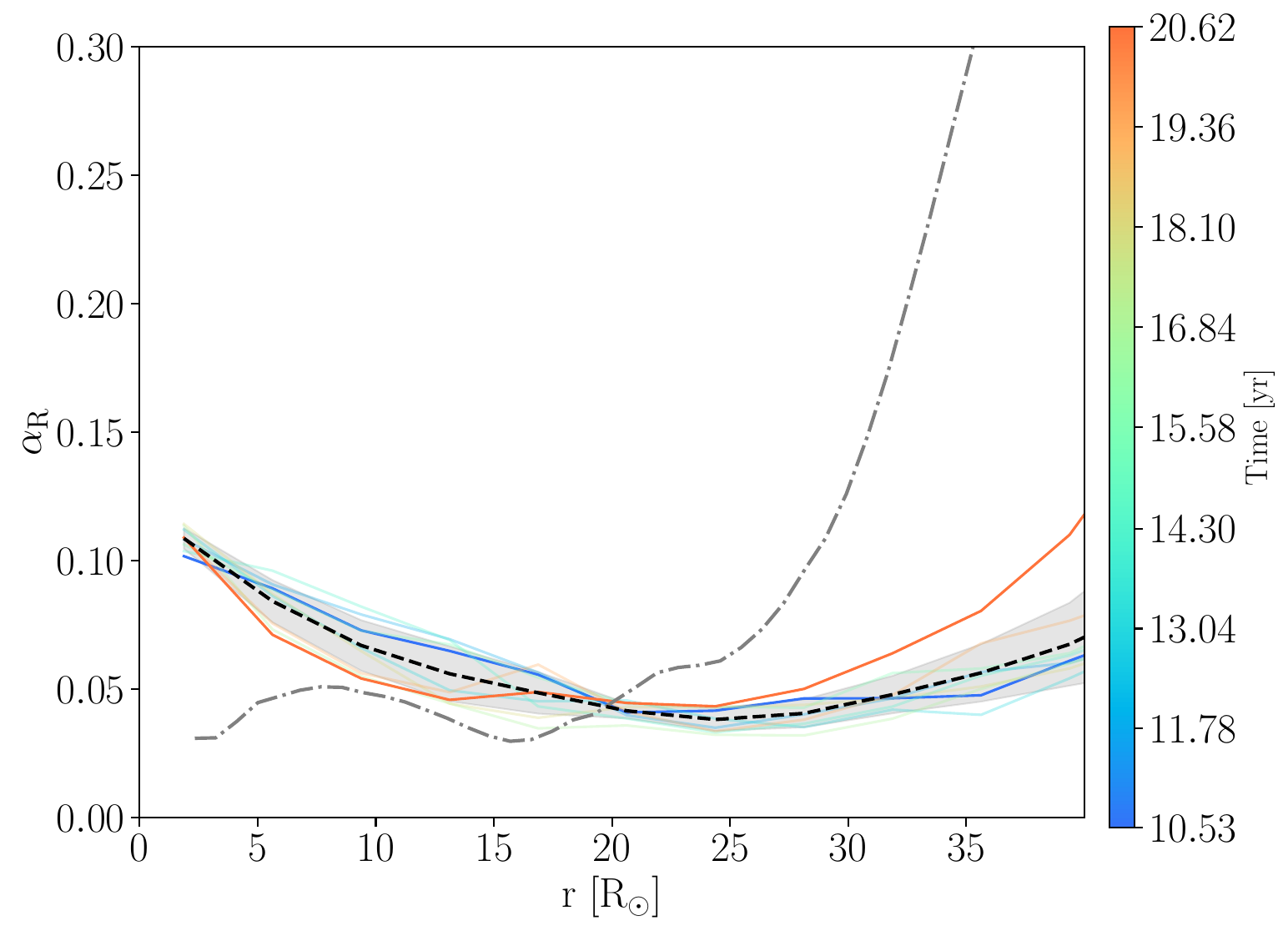}
    \caption{Time evolution of the short-term average profile ($\delta t=0.5$~yr) of the Reynolds stress parameter $\alpha_{\rm R}$, calculated with Eq.~\ref{eq:alpha}. Solid blue and orange lines denote the midpoint (10.5 yr) and final (21~yr) simulation stages, respectively. Intermediate profiles are rendered with transparency, following the colour gradient shown in the colour bar. The dashed black line represents the time-averaged $\alpha_{\rm R}$ over the $\Delta t =10$~yr, we represent the $1\sigma$ standard deviation across this period in grey. For comparison we added the time-averaged value of $\alpha_{\eff}$ as a dot-dashed gray line.}
    \label{fig:alpha_rey}
\end{figure}

In Figure~\ref{fig:alpha_rey}, we calculate the Reynolds stress to check if hydrodynamical turbulence could account for the observed level of accretion. The Reynolds stress captures the correlations between radial and azimuthal components of a turbulent velocity field with respect to a mean flow, and is defined as $\rho\langle\tilde{v}_r \tilde{v}_\phi\rangle$, where $\tilde{v}_r$ and $\tilde{v}_\phi$ are the turbulent fluctuations of the radial and azimuthal velocity components. Normalising by $\rho c_s^2$ and linking to the classic $\alpha-$prescription gives
\begin{equation}
    \alpha_{\rm R} = \frac{\langle \tilde{v}_r \tilde{v}_\phi\rangle}{c_s^2},
    \label{eq:alpha}
\end{equation}
where $\langle \tilde{v}_r \tilde{v}_\phi\rangle$ is calculated as follows. By dividing the disk into the same cylindrical shells as in Section~\ref{ssec:M_R_H}, for each shell we calculate the mean radial and azimuthal velocities by averaging over all SPH particles within the shell. The turbulent fluctuations for each particle are then obtained by subtracting the shell's mean from the particles velocities, i.e., $\tilde{v}_r = v_r - \langle v_r\rangle$ and $\tilde{v}_\phi = v_\phi - \langle v_\phi \rangle$. Finally, the product $\tilde{v}_r \tilde{v}_\phi$ is computed for each particle and averaged over the shell to obtain $\langle \tilde{v}_r \tilde{v}_\phi\rangle$. In \ref{app:alpha_and_am_loss} we demonstrate how this quantity is converged with respect to the shells' thickness.

Figure~\ref{fig:alpha_rey} shows the evolution of the short-term average ($\delta t~=~0.5$~yr) values of $\alpha_{\rm R}$ during the last 10 years of the disk's evolution. The radial profiles of $\alpha_{\rm R}$ present little temporal variation with values ranging 
between 0.04 and 0.11 at $r \leq 25$~\rsun. A comparison of $\alpha_{\rm R}$ and $\alpha_{\eff}$ (gray dot-dashed line) shows that, despite differences in their radial behaviour, both quantities remain consistent at the order-of-magnitude level across the disk.

This agreement suggests that hydrodynamical turbulence can account for the mass-transport rate observed in the simulation. Spiral shocks \citep{Hennebelle2017} are a possible additional source of angular momentum transport, although we do not find clear shocks in our disk at the current resolution, but this mechanism is discussed in Paper~I. The magnetorotational instability \citep{Balbus1991} is another angular momentum transport mechanism, but it requires magnetic fields that are not included in the current simulations. The only caveat is that our SPH simulations have not yet reached full numerical convergence. As shown in Figure~\ref{fig:app_mdot_alpaha_converge} (\ref{app:resolution_test}), the accretion rate during the first 0.5~yr is lower in the higher resolutions simulations. This behaviour suggests that numerical viscosity may contribute to the measured accretion rates and, consequently, to the inferred values of $\alpha_{\eff}$.

These values are broadly consistent with the 0.1 to 0.4 range reported in the literature for ionized disks in high mass X-ray transients \citep[e.g.,][]{King2007, Liu2007}. In contrast, 3D magnetohydrodynamics simulations typically report lower values ($\alpha \approx 0.001-0.05$) where Maxwell stress dominates angular momentum transport \citep{stone96,Hirose2006, Ju2017, Zhang2025}. The fact that our purely hydrodynamic simulations reproduce $\alpha$ values comparable to or larger than MHD results suggests that the hydrodynamical turbulence and spiral shock efficiently drive angular momentum transport even in the absence of magnetic fields. 

\subsection{Beyond the accretion disk: mass and angular momentum loss}
\label{ssec:turbulence_angular_momentum}

\begin{figure*}
    \centering
    \includegraphics[width=0.9\columnwidth]{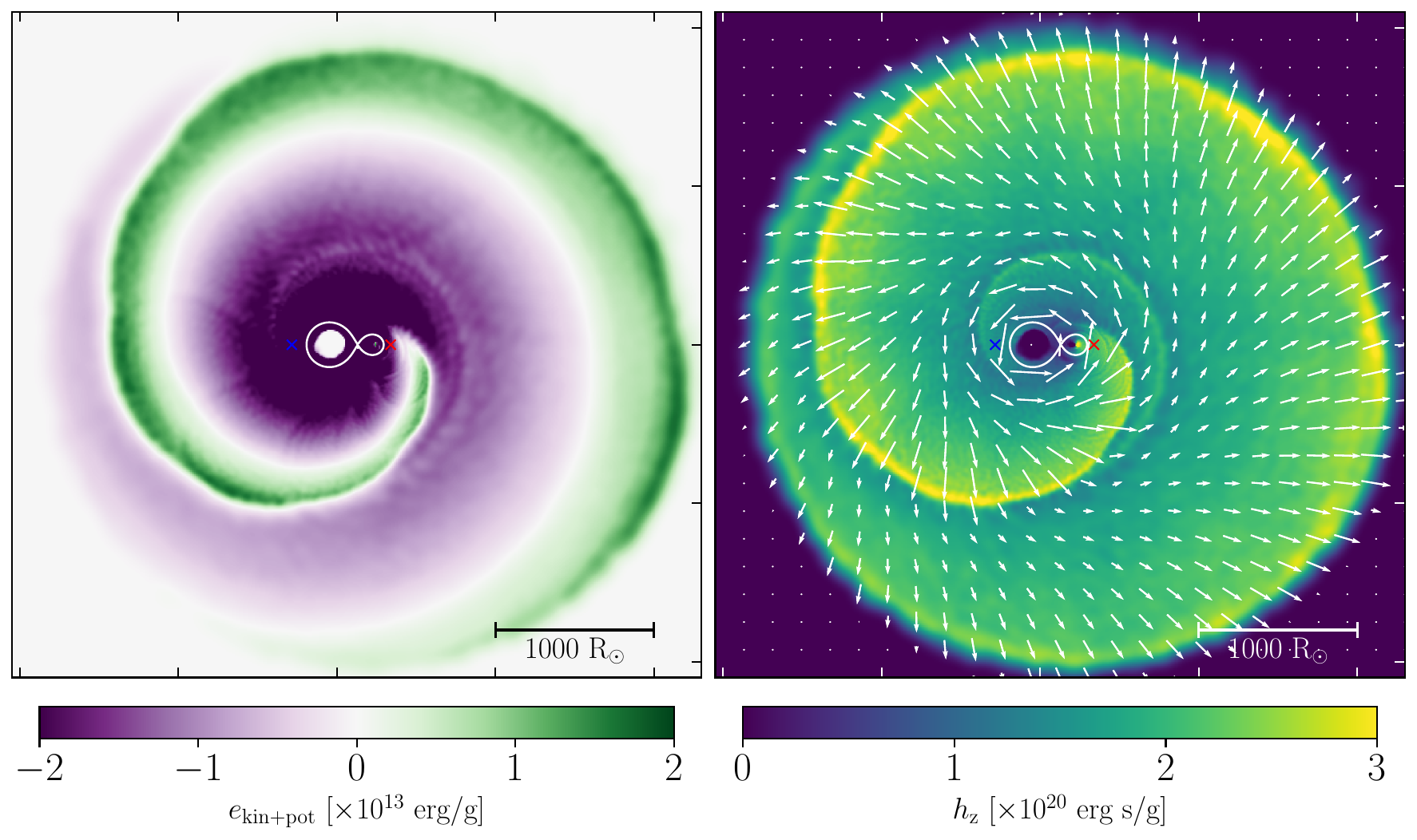}
    \caption{Left panel: cross-section of the specific orbital energy (kinetic + potential) in the orbital plane at 21~yr showing the whole simulation domain. Right panel: cross-section of the gas' specific angular momentum in the $z$-direction and velocity vectors in the orbital plane at 21~yr. The domain extends up to 2000~\rsun, everything outside of this radius is not part of the simulation. As reference we plotted the binary's Roche lobe contours in white, the position of the $L_{2}$ point (red cross), and the position of $L_{3}$ point (blue cross).}
    \label{fig:orbital_energy}
\end{figure*}

A key difference between this work and that of Paper~I lies in the size of the computational domain. While the domain in Paper~I was defined as 1.5 times the companion's Roche lobe radius ($124.65$~\rsun), the current simulation encompasses a much larger volume extending to a radius of 2000~\rsun\ from the \acrshort{COM}, which allows us to follow the evolution of outflows over a longer period. The left panel of Figure~\ref{fig:orbital_energy} shows a cross-section of the specific mechanical energy ($e_{\rm kin+pot}$) on the orbital plane across the entire domain at 21 yr. We observe a combination of bound gas ($e_{\rm kin+pot} <0$) and unbound gas ($e_{\rm kin+pot} > 0$) leaving the system from the vicinity of the second Lagrange point ($L_{2}$) and wrapping around the binary in a spiral pattern. This gas carries angular momentum and energy away from the system and may contribute to the formation of a circumbinary disk \citep[e.g.,][]{chen2017, Lu2023, Nibbs2025}.

The right panel of Figure~\ref{fig:orbital_energy} shows a slice of the $z$-component of the specific angular momentum on the orbital plane, $h_{\rm z}  = (\mathbf{r} \times \mathbf{v})_{\rm z}$, calculated with respect to the system's \acrshort{COM} in the inertial frame, where $\mathbf{r}$ and $\mathbf{v} = \mathbf{v}_{\rm sim}+\bm{\Omega}\times\mathbf{r}$ are the position and velocity of the gas in the inertial frame. The term $\bm{\Omega}\times\mathbf{r}$ represents the velocity transformation from the corotating frame ($\mathbf{v}_{\rm sim}$) to the inertial frame. The white arrows show the velocity field ($\mathbf{v}$) in the inertial frame. The mostly positive values of $h_z$ show the counter-clockwise rotation of the material. 

Specifically, the gas outflow originating from the vicinity of $L_{2}$ has high specific angular momentum and is associated with the unbound material shown in the left panel. Mass and angular momentum loss through the $L_{2}$ and $L_{3}$ points has been well studied \citep{Macleod2018, Reichardt2018, MacLeod2020, Scherbak2025}, with simulations showing that angular momentum is efficiently extracted from the binary, especially through the $L_{2}$ point.

Next we quantify the mass and angular momentum carried away by this outflowing gas and compare it with the literature. The mass outflow rate over a surface $S$ is given by
\begin{equation}
    \dot{M}_{\loss} = \int_{S} \rho\mathbf{v} \cdot d\mathbf{S},
    \label{eq:mass_loss}
\end{equation}
and the angular momentum loss rate by the gas crossing the same surface is defined as
\begin{equation}
    \dot{L}_{\loss, \rm z} = \int_{S}  h_{\rm z} \rho \mathbf{v} \cdot d\mathbf{S},
    \label{eq:am_loss}
\end{equation}
with $h_{\rm z}$ the gas' specific angular momentum in the $z$-direction, as previously defined. We focus on the $z$-component of the specific angular momentum because it aligns with the direction of the orbital axis. The specific angular momentum of the outflowing material is therefore
\begin{equation}
    h_{\loss} = \frac{\dot{L}_{\loss, \rm z}}{\dot{M}_{\loss}}.
    \label{eq:h_loss}
\end{equation}

\begin{figure}
    \centering
    \includegraphics[width=0.9\columnwidth]{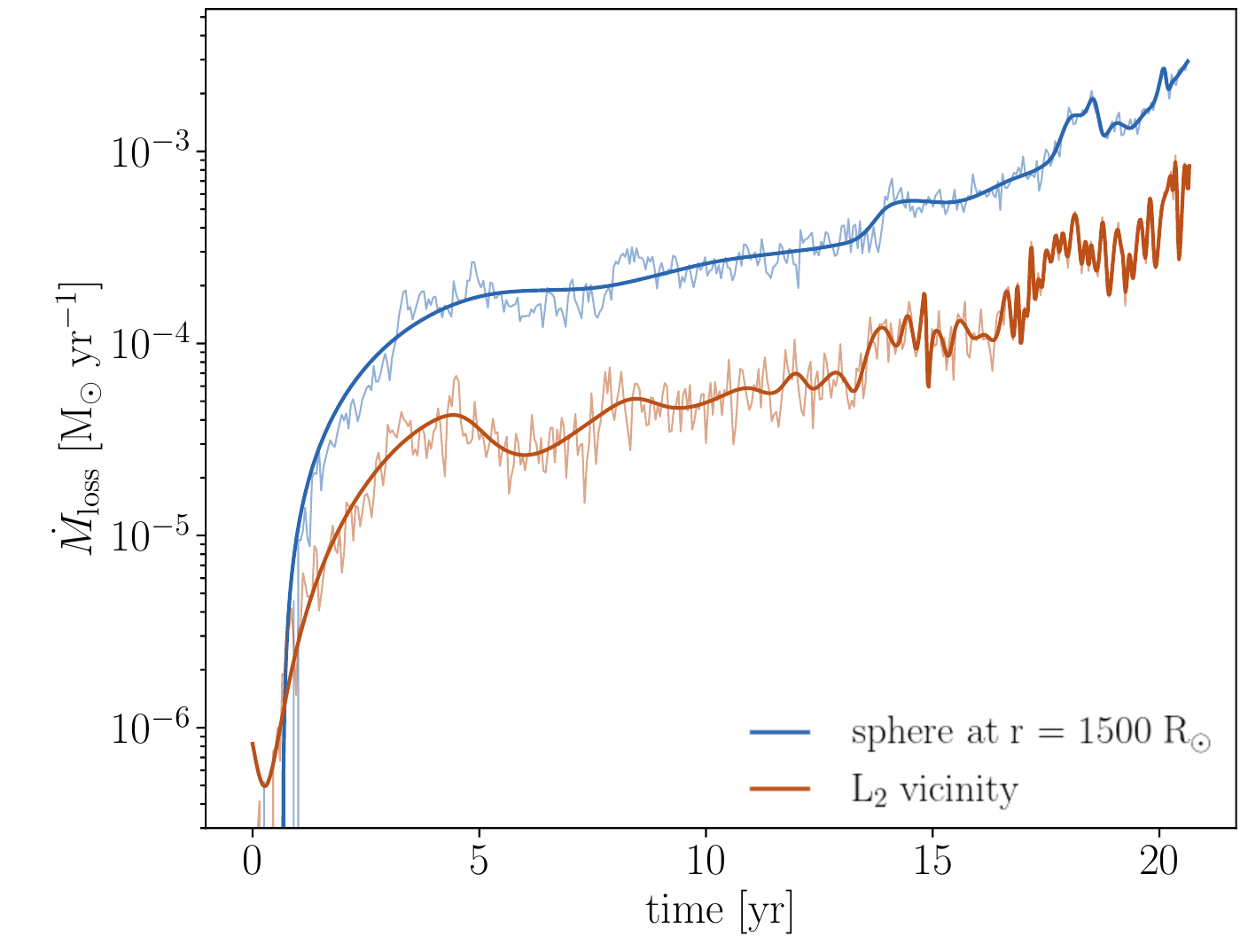}
    \caption{Mass-loss rate through the outer Lagrange point, $L_{\rm 2}$, versus time of binary outflows through a spherical shell centred on the binary's centre of mass (blue line), and through a cylindrical sector centred on $L_{\rm 2}$ (orange line). Solid lines indicate smoothed values.}
    \label{fig:mdot_loss}
\end{figure}

Note that this could also include energetically bound material. To estimate $h_{\loss}$, we define a spherical shell with a radius of 1500~\rsun\ (tests with 1000 and 2000~\rsun\ resulted in similar outcomes)
and a thickness $\Delta r = 1$~\rsun\ (see \ref{app:alpha_and_am_loss} for a convergence test) 
centred on the \acrshort{COM} and compute the flux of gas leaving the shell. Using Equations~\ref{eq:mass_loss}--\ref{eq:h_loss}, we measure a systemic mass outflow rate of $\dot{M}_{\loss}=2.9\times 10^{-3}$~\mdot\ by the end of the simulation (see Figure~\ref{fig:mdot_loss}, solid blue line) that carries angular momentum out of the system at a rate of 3.2\e{43}~cm$^2$ g s$^{-2}$. This estimate does not depend on the thickness and location of the spherical shell, see \ref{app:alpha_and_am_loss}. 

Next we quantify the mass leaving specifically through the surroundings of $L_{2}$. We follow the methodology of \citet[][see their section 3.2]{Scherbak2025}, although we define a control cylindrical sector around the outer Lagrangian point $L_{2}$, located at $r_{\Ltwo} = 339$~\rsun\ from the \acrshort{COM}, instead of a spherical cap. We use a radius of $1.1~r_{\Ltwo}$, a radial thickness of 1~\rsun,  extending between azimuthal angles $\phi = [-\pi/2, \pi/4]$, 
and with ${z} \in [0, 1.1~r_{\Ltwo}]$ (we later multiply the resulting mass outflow by two to extend the result to negative values of $z$). By using this cylindrical sector, positioned behind the $L_{2}$ point, we find a mass-loss rate of 8.7\e{-4}~\mdot\ (solid orange line in Figure~\ref{fig:mdot_loss}) carrying angular momentum at a rate of 7.0\e{42}~cm$^2$ g s$^{-1}$ (for a convergence test with respect to the thickness of the control surface see \ref{app:alpha_and_am_loss}). To compare the two methods, we compute the specific angular momentum of the outflow $h_{\loss}$ (Equation~\ref{eq:h_loss}); we find that $h_{\loss}$ from the cylindrical sector is 72\% of the value obtained from the full sphere. This confirms that the $L_{2}$ outflow accounts for the majority of the system's mass and angular momentum loss.  

\begin{figure}
    \centering
    \includegraphics[width=\columnwidth]{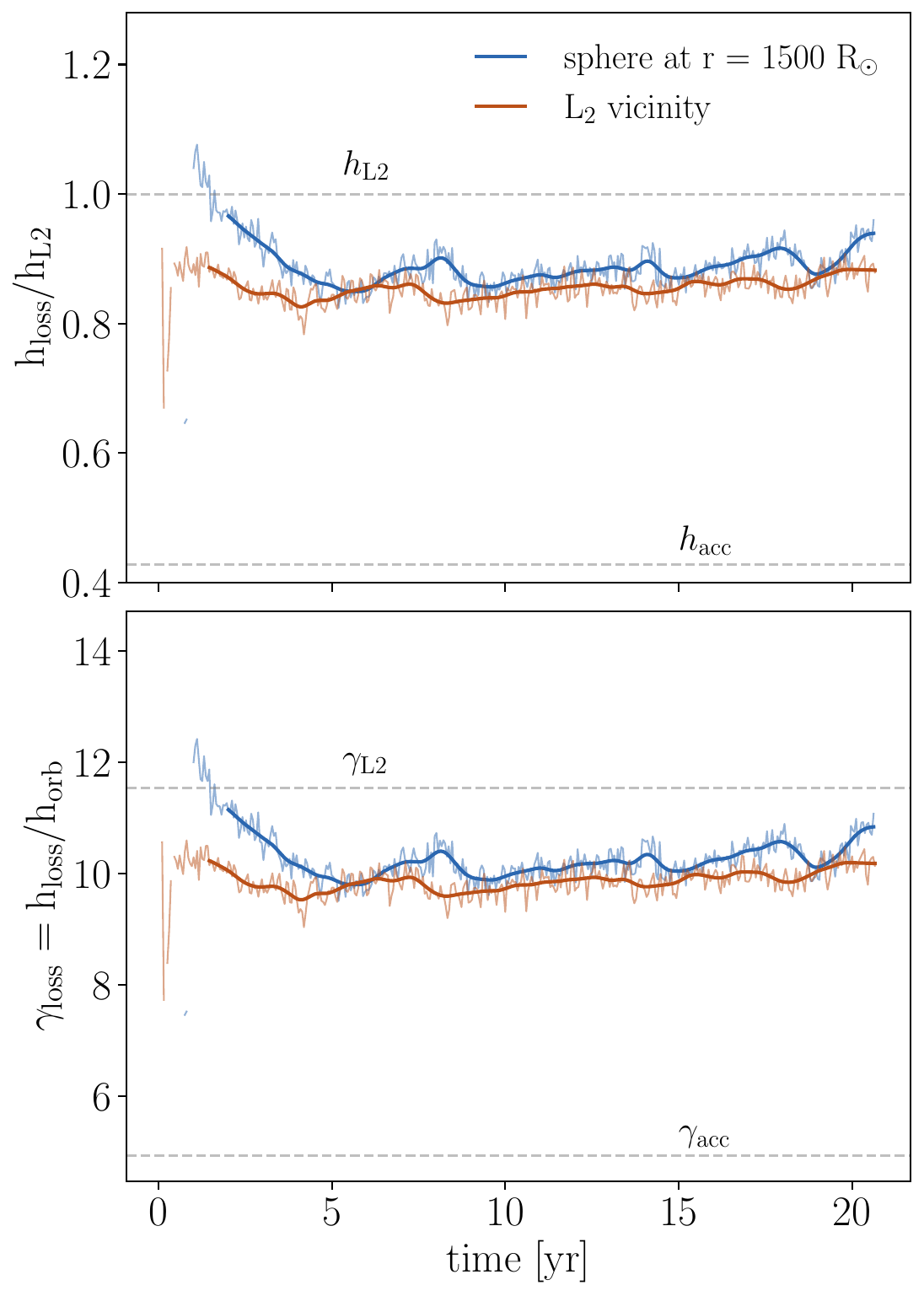}
    \caption{Top panel: average specific angular momentum of the outflowing material as a function of time, normalised by the specific angular momentum of $L_{2}$ ($h_{\Ltwo}$). Bottom panel: the specific angular momentum of the outflowing material normalised by the specific angular momentum of the binary ($h_{\orb}$). Solid lines represent smoothed simulation results. The dashed gray lines represent reference values normalised by $h_{\Ltwo}$ (top panel) and $h_{\orb}$ (bottom panel): $h_{\rm acc}$ and $\gamma_{\rm acc}$ represent the accretor's specific angular momentum, while $h_{\Ltwo}$ and $\gamma_{\Ltwo}$ represent the $L_{\rm 2}$ specific angular momentum. Details on the calculation of these quantities are given in the text.}
    \label{fig:am_loss}
\end{figure}

In the top panel of Figure~\ref{fig:am_loss}, we show the evolution of $h_{\loss}$, normalised by the specific angular momentum at the $L_{2}$ point ($h_{\Ltwo}  = \Omega r^2_\mathrm{L2}$). The angular momentum loss of both the outflowing material and the material specifically escaping through the $L_{2}$ region reach similar values relative to $h_{\Ltwo}$ with $h_{\loss}\approx 0.93~h_{L2}$ for the sphere method (blue line), and $h_{\loss} \approx 0.87~h_{L2}$ for the flow estimated in $L_{2}$ vicinity (orange line). This value remains nearly constant throughout the 21~yr simulation, except at the beginning of the simulation where the mass-loss rate is negligible. The value of $h_{\loss}$ is close to the upper limit, $h_{L2}$.
These findings are consistent with the results from \cite{Macleod2018} and \cite{Macleod2020b}, who found the specific angular momentum loss to be between that of the accretor ($h_{\rm acc} ~=~\Omega r_{\rm acc}^2$) and that of the $L_{2}$ point. \cite{Scherbak2025} using 3D simulations of stable mass transfer ($\dot{M} \gtrsim 10^{-4}$~\mdot) found that the ratio $h_{\loss}/h_{\Ltwo}$ decreases when the mass ratio increases ($q$), specifically for binary system with $q=0.25$, they measured an angular momentum loss of $h_{\loss} \approx 0.95~h_{\Ltwo}$, similar to our estimate.

We also compare $\gamma_{\loss} =h_{\loss}/h_{\orb}$,  where
\begin{equation}
    h_{\orb}= \frac{M_dM_a}{ M_{\rm tot}^2} \sqrt{GM_{\rm tot}a}, 
\end{equation}
is the specific angular momentum of the binary for a circular orbit ($e$~=~0), with $M_{\rm tot}= M_d + M_a$. The ratio $\gamma_{\loss}$ is a critical parameter in binary evolution, as it controls the rate of orbital shrinkage, 
and hence how quickly the system moves into a non-conservative regime \cite[see equation~2 in][]{Macleod2018}. In our simulation, the mass escaping the system carries on average 10.7 times more specific angular momentum than the binary orbit. These values of $\gamma_{\loss}$ are consistent with those found by \cite{Macleod2020b}, who performed 3D hydrodynamic simulations of coalescing binary systems, with mass ratios similar to ours ($q= M_{a}/M_{d} = 0.1, 0.3$), and rapid mass transfer (with a range of $\dot{M}$~=~1\e{-2} -- 1\e{-1}~\mdot\ by the end of their simulations), and reported values of $\gamma_{\loss} \geq 8$. We also show in Figure~\ref{fig:am_loss} that $\gamma_{\loss} \sim \gamma_{L2}$, which aligns with the theoretical predictions from mass loss originating from the $L_{2}$ point \citep{Pribulla1998, Pejcha2014, Pejcha2017}. This suggests that the mass loss process is governed by particles moving through $L_{2}$ and forming an equatorial outflow, instead of leaving the binary via an isotropic wind from the accretor.

Finally, we compare the orbital decay predicted by the 1D {\sc mesa} model, which adopts a fully conservative mass transfer prescription (Section~\ref{sec:setup_1D}), and the orbital decay predicted by our measurement of $L_{2}$ mass loss. We solve for the orbital separation evolution for non-conservative mass transfer, \citep{Huang1963, Macleod2018, Macleod2020b},  

\begin{equation}
    \frac{\dot{a}}{a} = -2 \frac{\dot{M}_{\rm d}}{M_{\rm d}}\left[1- \beta\left(\frac{M_{\rm d}}{M_{\rm a}}\right)-(1-\beta)\left(\gamma_{\rm loss} + \frac{1}{2}\right)\frac{M_{\rm d}}{M_{\rm d} + M_{\rm a}}\right], 
    \label{eq:orbital_evolution}
\end{equation}

\noindent where $\beta = \dot{M}_{\rm acc}/\dot{M}_{\Lone}$ is the ratio between the measured mass accretion rate onto the companion ($\dot{M}_{\rm acc}$), and the mass transfer rate from \mesa\ ($\dot{M}_{\Lone}$), and $\gamma_{\rm loss}$ is our measured value of the specific angular momentum of the escaping material. 

\begin{figure}
     \centering
     \includegraphics[width=0.85\columnwidth]{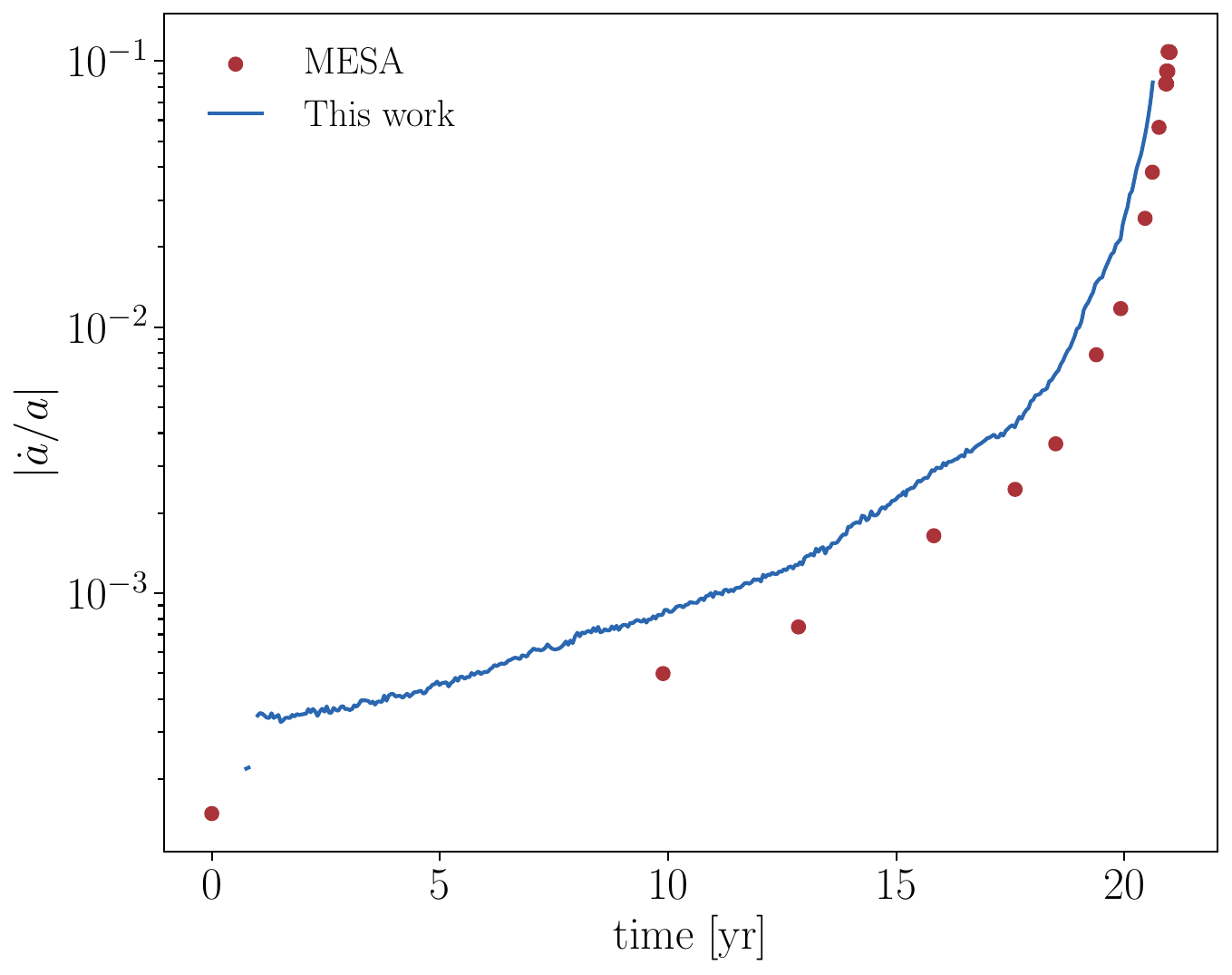}
     \caption{Orbital decay over the 21 years of the simulation, reproduced measured with our values of mass accretion rate in Figure~\ref{fig:mdot_simulation}, and $\gamma_{\loss}$ values in Figure~\ref{fig:am_loss}. We show the values prescribed by the \mesa\ simulation (red dots) for comparison.}
     \label{fig:inspiral-timescale}
\end{figure}

Figure~\ref{fig:inspiral-timescale} shows the orbital decay in these two models in terms of the inverse timescale parameter $|\dot{a}/a|$. The discrepancy between the two curves shows non-conservative mass transfer leads to a faster inspiral in our simulation compared to the \mesa\ simulation, by a factor of $\sim$2 at 20.5~yr. The timescale estimated from the SPH simulation at $t=1.2$~yr is formally shorter than that prescribed by {\sc mesa} but, only a handful of particles have reached $L_{2}$ at this stage, making the estimate uncertain.

Extrapolating the orbital decay beyond the 3D simulated evolution, assuming the mass transfer rate and the value of $\gamma_{\loss}$ remain fixed at their last simulated values, we find that the binary separation reaches the donor's radius ($R_{\rm don} = 144.8$~\rsun) after an additional 6~yr, i.e. at $t=27$~yr. This marks an upper limit for the onset of the CE phase, since at the end of our simulations the mass transfer rate keeps growing with time.

\section{Effect of physical and numerical parameters}
\label{sec:differences}

In this section, we assess how our assumptions (Table~\ref{tab:simulations}) affect the properties of the accretion disk. We start with a comparison of the SPH and grid-based results using a simulation designed to closely match that of Paper~I. We then examine the effects of the nozzle size (Section~\ref{ssec:comp_nozz}), and the 3D simulation length (Section~\ref{ssec:comp_time}), and the adopted gas adiabatic index (Section~\ref{sec:comp_gamma}).

\subsection{Comparison with Paper~I}
\label{ssec:comp_codes}


Here, we compare the simulation carried out with the grid-based \mezcal\ adaptive mesh refinement
({\it sim-0} in Paper I) to the results of this paper. 
For this purpose we have carried out a new \phant\ SPH simulation, called {\it sim-compare} in Table~\ref{tab:simulations}, with input parameters that are as close as possible to the simulation of Paper~I, with none of the simulation improvements implemented in our fiducial simulation, {\it sim-fiducial}.

\begin{figure}
    \centering
    \includegraphics[width=\columnwidth]{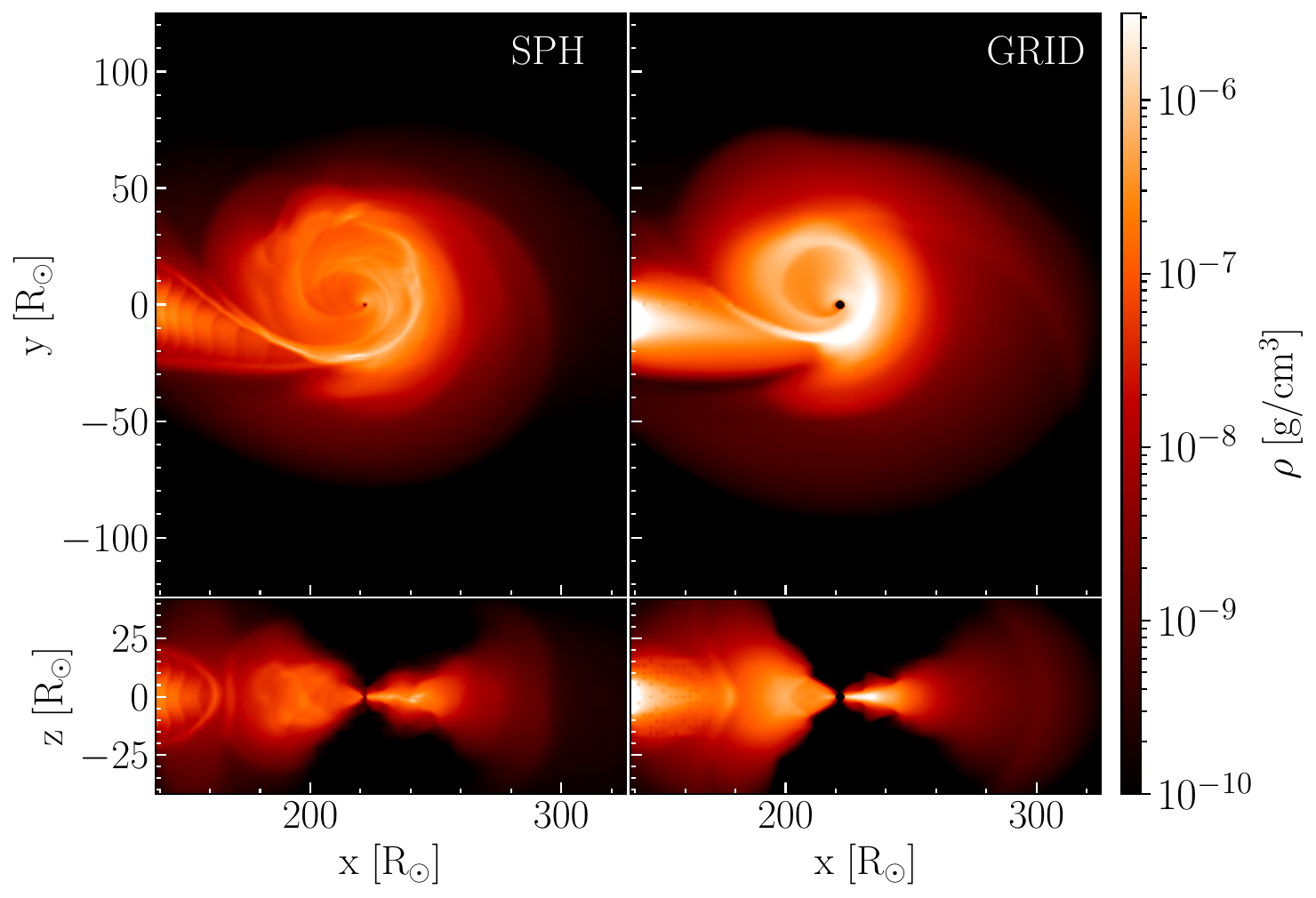}
    \caption{Density cross-sections in the orbital plane (top panels) and perpendicular plane (bottom panels) of the accretion disk at 21 yr, for our controlled simulation ({\it sim-compare} in Table~\ref{tab:simulations}) and for the grid-based simulation of Paper~I ({\it sim-0} in that paper).}
    \label{fig:grid_vs_sph}
\end{figure}    
Density slices at 21~yr for the two simulations are compared in Figure~\ref{fig:grid_vs_sph}. The \mezcal\ spatial resolution ranges between $\Delta x = \Delta y = \Delta z = 2.64~R_\odot$ (1.85\e{11}~cm) and $\Delta x = \Delta y = \Delta z = 0.65~R_\odot$ (4.53\e{10}~cm). The  \phant\ simulation ({\it sim-compare}) adopts a particle mass of $1 \times 10^{-10}$~\msun, resulting in 1.86 million gas particles by 21 years.
Within the companion's Roche lobe, the SPH particles' smoothing lengths range from $h=0.18$~\rsun\ to $9.23$~\rsun. The grid simulation therefore has higher resolution than SPH in the lowest density regions, whereas the SPH simulation reaches roughly three-times higher resolution than the grid in the denser regions. 

Both simulations produce an accretion disk with some differences in the density structure. In the SPH simulation ({\it sim-compare}), gas particles are injected in discrete layers as is visible in Figure~\ref{fig:grid_vs_sph} (left column). The stream is poorly resolved because it is stretched as it falls into the accretor Roche lobe, making individual layers of particles visible as density perturbations in the slice view. The grid simulation instead injects material continuously across the nozzle, producing a smooth, denser, and well-resolved stream towards the accretor. A striking visual difference between the two simulations are additional substructures in the SPH disk. This turbulent substructure also becomes more apparent as the numerical resolution is increased (Figure~\ref{fig:apx_renders}).


\begin{figure}[ht!]
    \centering
    \includegraphics[width=0.8\columnwidth]{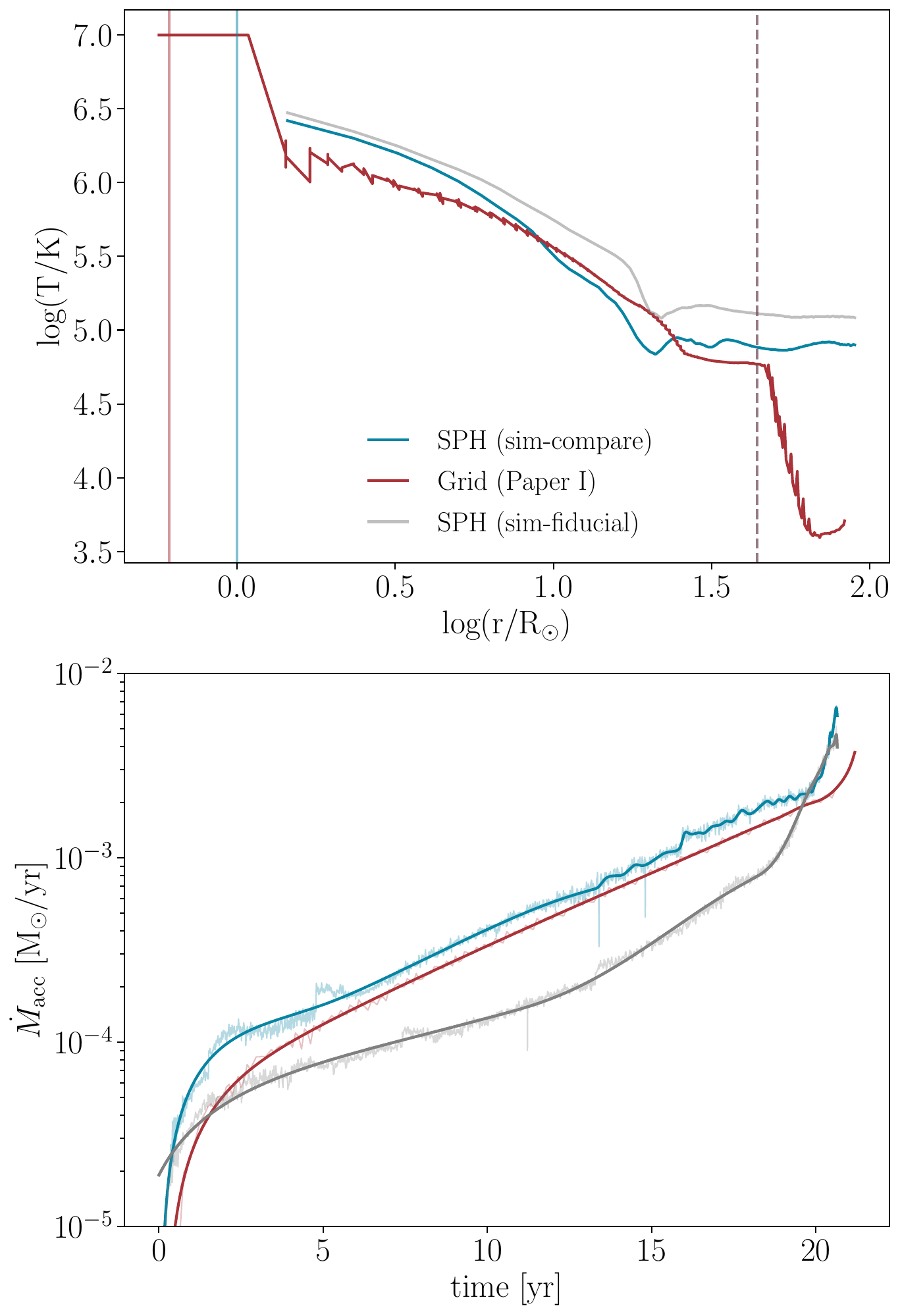}
    \caption{
    A comparison of the temperature profiles at 21~yr (top panel) and of the evolution of the mass accretion rate onto the companion (bottom panel) for the grid simulation of Paper~I (red solid curve), the SPH simulation calculated as a direct comparison (blue solid line) and the SPH simulation with the improvements developed in this work (grey solid line). The inner disk boundaries in the grid and SPH simulations are marked by the red and blue vertical solid lines, respectively, while the dashed vertical line shows the disk outer radius.}
    \label{fig:grid_vs_sph_mdot_temp}
\end{figure}    

We compare the mass, radius and scale height of the disk 
in Table~\ref{tab:disk_parameters}. As expected, the properties of the comparison SPH simulation ({\it sim-compare}) show significantly better agreement with the grid results than the fiducial simulation does. 
That said, disk masses and scale heights are within 10\% of one another, and the disk radii are almost unchanged.
Figure~\ref{fig:grid_vs_sph_mdot_temp} (top panel) shows the radial temperature profile for the grid simulation of Paper I (red line), the control simulation (blue line), and fiducial simulation (grey line). They agree within 10\%, although the discrepancy increases to $\sim$25\% beyond the measured outer disk's radius.

The accretion rate evolution (Figure~\ref{fig:grid_vs_sph_mdot_temp}, bottom panel) is similar for the grid and SPH simulations, with an offset of about 1 year. The final values of the mass accretion rate are 6.0\e{-3}~\mdot (SPH) and 3.7\e{-3}~\mdot\ (grid), with the SPH accretion rate being on average 24\% larger throughout the simulations. Given the differences in numerical methods, spatial resolution, and measurement of accretion between the two codes, a 24\% discrepancy in the accretion rate is small. 
Further, differences in the two profiles may be expected due to the impact of numerical viscosity which is likely different between the two simulations: while the grid simulation produces a fully converged accretion rate (figure~16 in Paper~I), the SPH simulations do not exhibit full convergence across three particle-mass resolutions considered (Figure~\ref{fig:app_mdot_alpaha_converge} in \ref{app:resolution_test}).

We conclude that the disk simulated with the two methods (grid-based and SPH) exhibits quantitatively similar properties. 
In the next Section we investigate the effect of the setup improvements in detail.

\subsection{The impact of injection parameter choices}
\label{ssec:comp_nozz}

The main differences between the current SPH and grid-based injection schemes are: (1) the adoption of the optically thick mass-transfer prescription of \citet{Kolb90} instead of \citet{Ritter1988}, resulting in an increasing nozzle radius, $S_{\Lone}$; and (2) a supersonic injection velocity in the SPH simulations, compared to a subsonic velocity in the grid-based model. The optically thick mass transfer regime of \cite{Kolb90},  which implies an increasing nozzle radius ($S_{\Lone}$) during \acrlong{RLOF} is used for the SPH simulation instead of the prescription of \cite{Ritter1988}, which is not grossly different during the first $\sim15$~yr of the simulation (see Figure~\ref{fig:app_rad_nozz} in \ref{app:appendix_radius}). Moreover, in Paper~I, the fluid was injected with subsonic velocity ($\mathcal{M}_{\Lone}~=$~0.1), while in this paper, we inject gas particles with sonic velocity ($\mathcal{M}_{\Lone}~=$~1.0), which is more physically motivated \citep{lubow1975gas}.

To isolate the effects of these two modifications, we performed two additional simulations. (i) First, we adopted the optically thick mass-transfer prescription to determine the nozzle radius, {\it but} retained the lower injection velocity of Paper~I ($\mathcal{M}_{\Lone}=0.1$). (ii) Next, we used the nozzle radius from Paper~I (based on the optically thin prescription) {\it but} adopted a higher injection velocity, $\mathcal{M}_{\Lone}=1.0$. These models are called {\it sim-rad-nozzle} and {\it sim-mach-01}, respectively (see Table~\ref{tab:simulations}).

Based on the global properties listed in Table~\ref{tab:disk_parameters}, we find that, at 21~yr, \textit{sim-mach-01} differs from \textit{sim-fiducial} primarily in its disk scale height, which is 60\% larger, while all other quantities agree to within 15\%. By comparison, \textit{sim-rad-nozzle} remains within 5\% of the fiducial model for all quantities except the disk mass and accretion rate, which differ by 20\% and 15\%, respectively. These results indicate that the simulations are more sensitive to the injection velocity than to the nozzle radius.

\begin{figure}
    \centering
    \includegraphics[width=0.85\columnwidth]{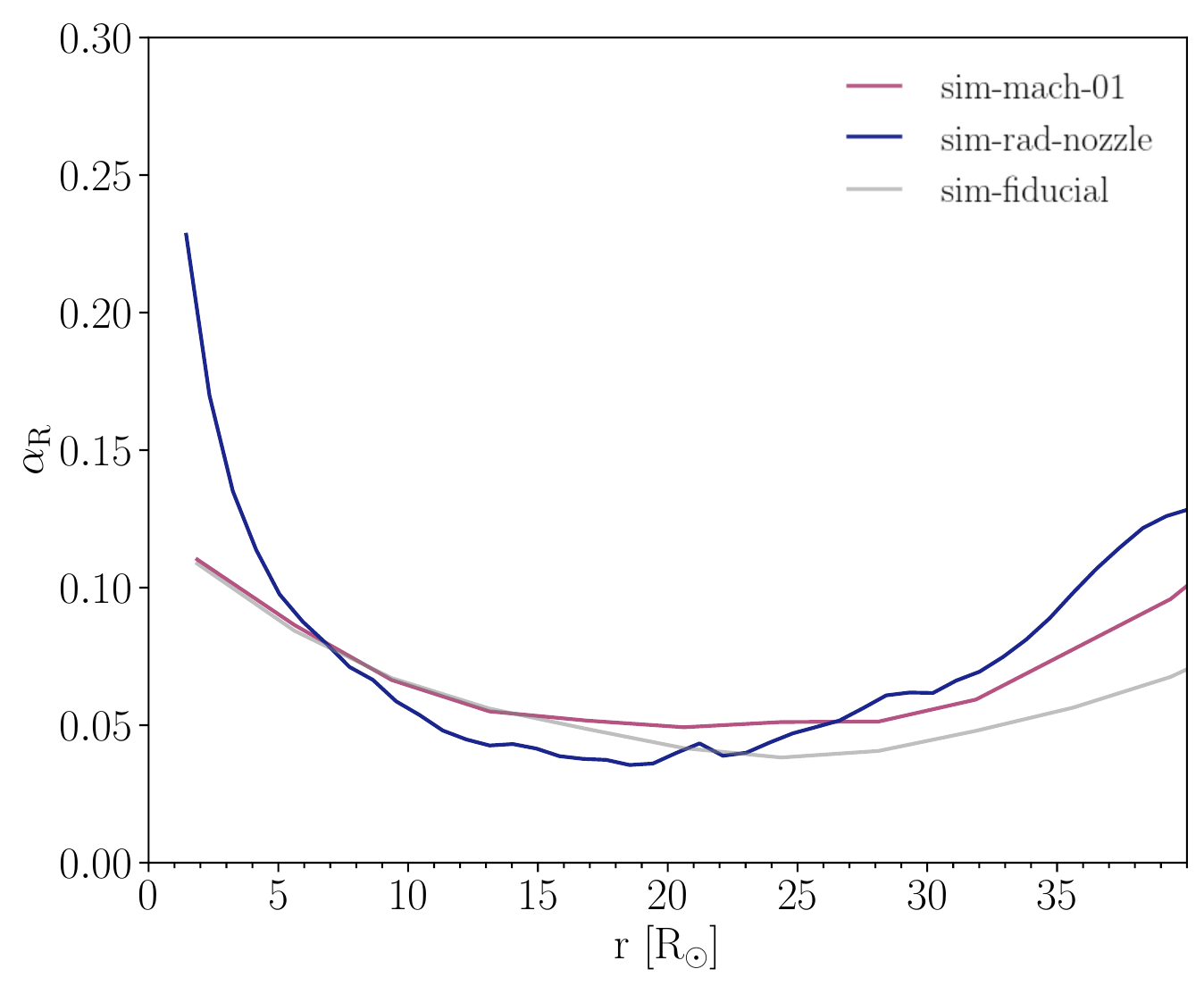}
    \caption{Short-term average profiles ($\delta t =0.5~yr$) of the viscosity parameter measured with the Reynolds stress method, $\alpha_{\rm R}$, for simulations {\it sim-mach-01} (magenta line), {\it sim-rad-nozzle} (blue line) compared to {\it sim-fiducial} (gray line) measured at $t~=~21$~yr.}
    \label{fig:alpha_nozzle}
\end{figure}    

Figure~\ref{fig:alpha_nozzle} shows the short-term averages ($\delta t =0.5$~yr) radial profile of $\alpha_{\rm R}$ at $t=21$~yr for the three simulations {\it sim-fiducial}, {\it sim-mach-01} and {\it sim-rad-nozzle}. Adopting a constant nozzle radius (blue line) increases $\alpha_{\rm R}$ in the inner disk, reaching values up to a factor of two larger at the accretion radius ($h_a = 1$~\rsun). In contrast, changing the injection velocity has only a modest effect on the $\alpha_{\rm R}$ profile within the disk circularization radius, while larger differences appear in the outer disk, where the slower injected particles accelerate. Overall, the three profiles agree to within $\sim$30\% and remain in the range $\alpha_{\rm R}\simeq0.02$--0.12 over $5 \lesssim r \lesssim 25$~\rsun. This agreement indicates that the level of turbulence within the disk is only weakly sensitive to the adopted mass-injection prescription.

\begin{table*}[ht!]
    \centering
    \begin{tabular}{lc|cccc|cccc}
    \hline
    &&\multicolumn{4}{c|}{Input parameters} &\multicolumn{4}{c}{Measured parameters}\\
    \hline
        Model & 3D simulation &   Donor &    Companion &Binary &$\dot{M}_{\ini}$&Disk&Disk & Disk Scale & Mass accretion \\
 & time& mass& mass& separation&  & Mass& Radius& Height &  rate  \\
 & [yr]&  [\msun]& [\msun]& [\rsun]& [\mdot]& [\msun]& [\rsun]& [\rsun] & [\mdot]  \\
        \hline 
        sim-10yr&  10&  6.96&    1.42&263&1.5\e{-3}& 1.1\e{-2}& 39& 5.0 & 2.0\e{-3} \\
       sim-fiducial&  21& 6.98&    1.40&267&1.3\e{-4}&5.0\e{-3}&40& 5.0 & 5.2\e{-3} \\
        sim-45yr&  45&  6.97& 1.41& 267& 5.8\e{-5} & 4.6\e{-3}& 45 & 7.0 & 1.3\e{-2} \\

 \hline
 \end{tabular}
     \caption{Input parameters and measured properties for different simulation starting points. The 3D simulation start time indicates the time remaining before the end of the mass transfer rate, and $\dot{M}_{\ini}$ is the initial mass transfer rate calculated by \mesa. The value of the mass transfer rate for all simulations at the end of the simulation is $\dot{M}_{\fin}~=~1.03\times 10^{-1}$~\msun~yr$^{-1}$. All properties (disk mass, radius, scale height) and mass accretion rate onto the companion are estimated at the end of the simulation.}
   
    \label{tab:dif_times}
\end{table*}

\subsection{Effect of neglecting the preceding evolution}
\label{ssec:comp_time}

All our 3D simulations follow the final 21 years of the 30\,000~yr mass transfer phase modelled in 1D (Section~\ref{sec:setup_1D} and Figure~\ref{fig:mdot_mesa}), when the mass transfer rate increases exponentially. To assess the impact of the adopted starting point along the mass-transfer sequence, we performed two additional simulations spanning 10 and 45~yr, respectively. Input and measured parameters of the three simulations are presented in Table~\ref{tab:dif_times}.

The viscous timescale of the accretion disk $t_{\rm visc}\sim R_{\disk}^2/\nu\approx$~10~yr is estimated from the disk properties at 21~yr using $\nu = \alpha c_s H_{\disk}$, with $\alpha =0.1$ corresponding to the value of $\alpha_{R}$ at the inner radius, $r= h_{a}$). This timescale represents the characteristic relaxation time of the disk and thus sets the timescale over which the influence of the initial conditions is expected to diminish. For a constant mass injection rate, the disk is expected to reach a steady state only if the viscous timescale is shorter than the mass injection timescale. In our simulations, the modeled durations (10, 21, and 45~yr) are all comparable to the estimated viscous timescale of the disk.

\begin{figure*}[ht]
\centering
\includegraphics[width=\columnwidth]{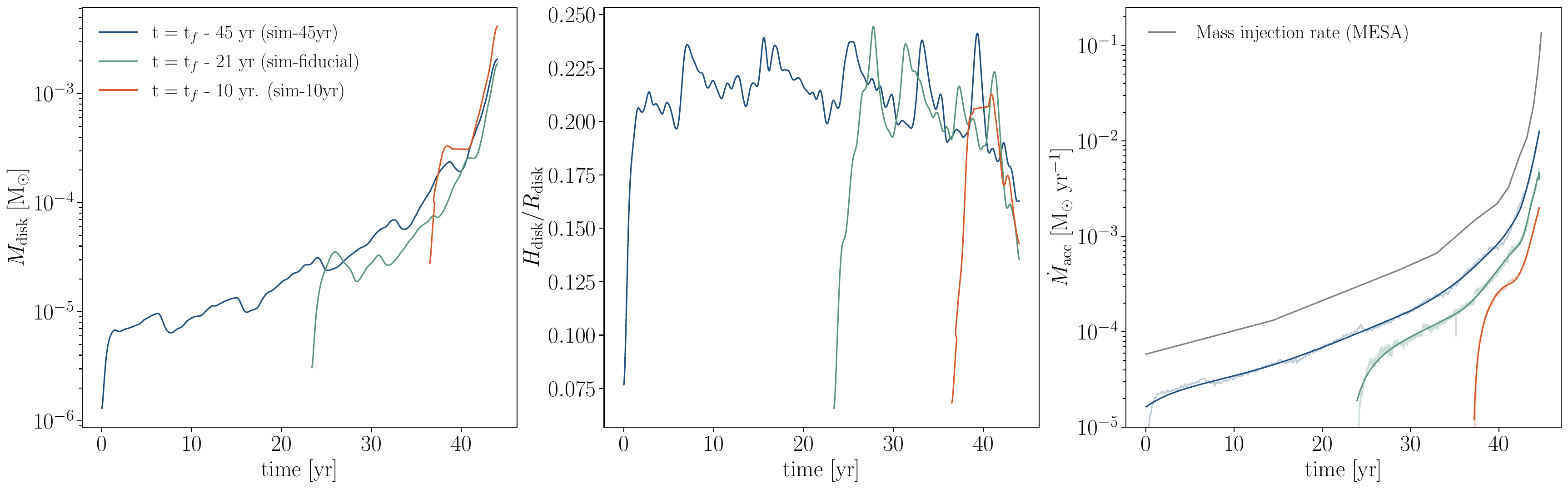}
\caption{Evolution of the disk mass (left panel), aspect ratio (middle panel), and mass accretion rate ($\dot{M}_{\rm acc}$, right panel) for the three simulations that were run for different time lengths (10, 21, and 45 years). The gray line in the right panel represents the mass injection rate given by the 1D \mesa\ simulation.}
\label{fig:disk_evolution}
\end{figure*}    

Figure~\ref{fig:disk_evolution} shows the evolution of the disk mass, aspect ratio ($H_{\disk}/R_{\disk}$), and accretion rate for the three simulations. Overall the disk mass exhibits the largest variations among the simulations, reaching a factor of two. This difference is reflected in the mass accretion rate evolution and results from the smaller amount of mass injected in simulations started at later times. On the other hand, the aspect ratio differs by less than $\sim$10\%, indicating that the structural properties of the disk are robustly captured across the different start times.

The mass accretion rate onto the companion (Figure~\ref{fig:disk_evolution}, right panel) increases by several orders of magnitude in all three simulations on a timescale shorter than the disk's viscous timescale ($t_{\rm visc}\sim 10$~yr). Consequently, the accretion flow cannot be assumed to have reached a steady state. This effect is most pronounced in the 10 year simulation (orange line), whose duration is comparable to the viscous time and which ends with an accretion rate nearly an order of magnitude lower than that of the 45~yr simulation.

\begin{figure}[ht]
    
    \includegraphics[width=0.8\columnwidth]{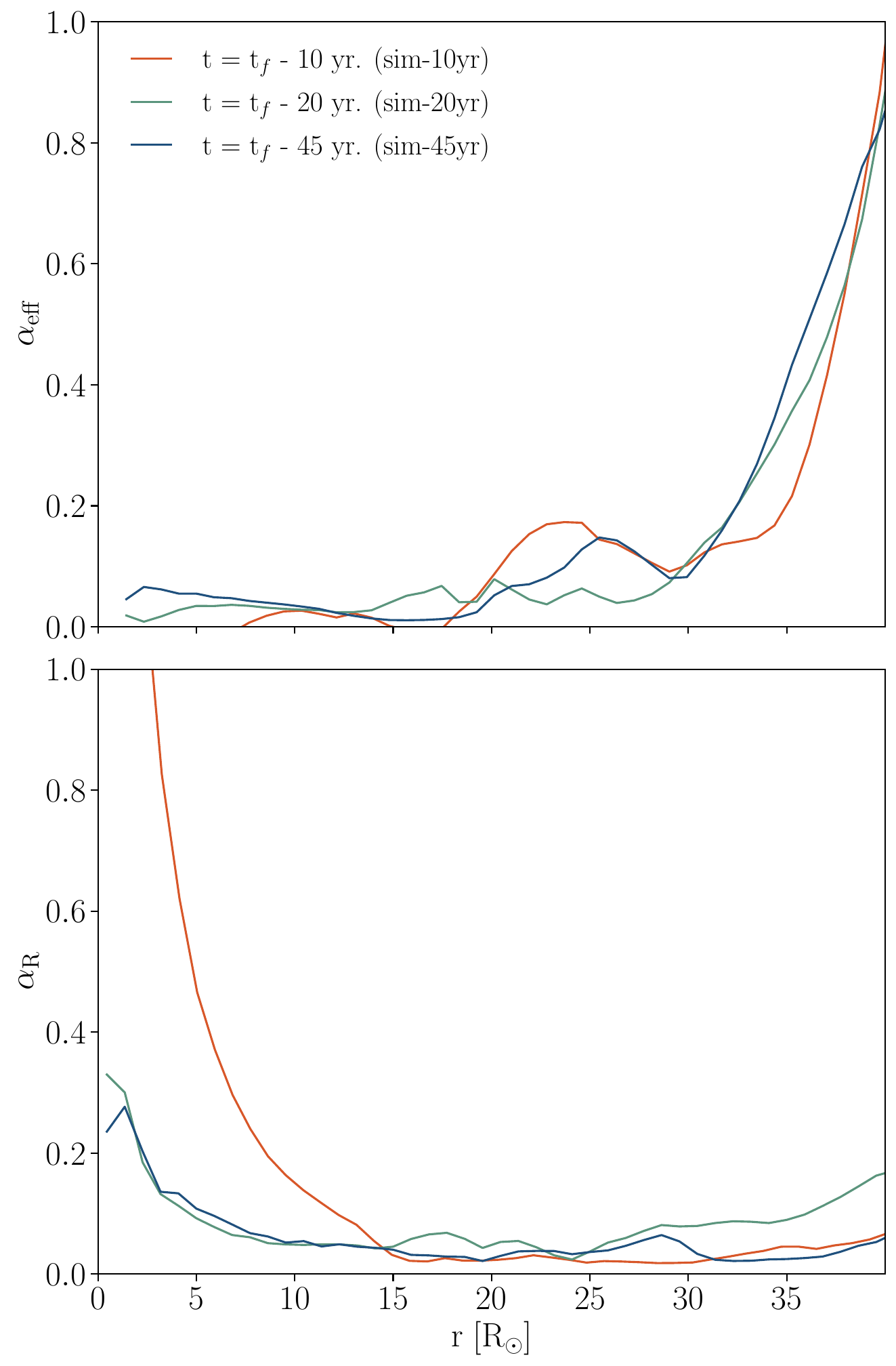}
    \caption{Radial profiles of $\alpha_{\eff}$ and $\alpha_{\rm R}$ at the end of simulations with different starting times. The two quantities are calculated using Equations~\ref{eq:alpha_ss} and \ref{eq:alpha}, respectively.}
    \label{fig:alpha_long}
\end{figure}    

Finally, we compare the profiles of $\alpha_{\eff}$ and $\alpha_{\rm R}$ at the end of the three simulations (Figure~\ref{fig:alpha_long}). The $\alpha_{\eff}$ profiles are very similar, differing only through local fluctuations that are smaller than the intrinsic temporal variability of each simulation. For $\alpha_{\rm R}$ (bottom panel), noticeable differences appear below $\sim 15$~\rsun, where the 10~yr simulation (orange curve) exhibits systematically larger values.

Overall, varying the starting point along the mass-transfer sequence primarily affects the total mass supplied to the disk and, consequently, the disk mass and accretion rate. In contrast, the disk structure, as characterized by its aspect ratio and the radial profiles of $\alpha_{\eff}$ and $\alpha_{\rm R}$, remains broadly unchanged. These results indicate that our main conclusions regarding disk properties and angular momentum transport are robust to the adopted initial conditions.

\subsection{The choice of adiabatic index}
\label{sec:comp_gamma}
During the mass-transfer phase, shocks and spiral arms efficiently compress and heat the gas. The resulting pressure can prevent disk formation unless the generated thermal energy is radiated away. For an ideal gas equation of state (Eq.~\ref{eq:idealgas}), the adiabatic index $\gamma$ determines the thermodynamic response of the gas to compression: larger values of $\gamma$ produce a stronger pressure increase for a given density enhancement, reducing compressibility and making it more difficult for the gas to settle into a rotationally supported disk. Adoption of a low adiabatic index close to the isothermal limit of $\gamma = 1$, is commonly used as a numerical proxy for efficient cooling. Following the methodology of Paper~I, we performed new simulations with different values of the adiabatic index (Table~\ref{tab:simulations}) and compared their outcome below.

\begin{figure*}[ht]
    \centering
    \includegraphics[width=\textwidth]{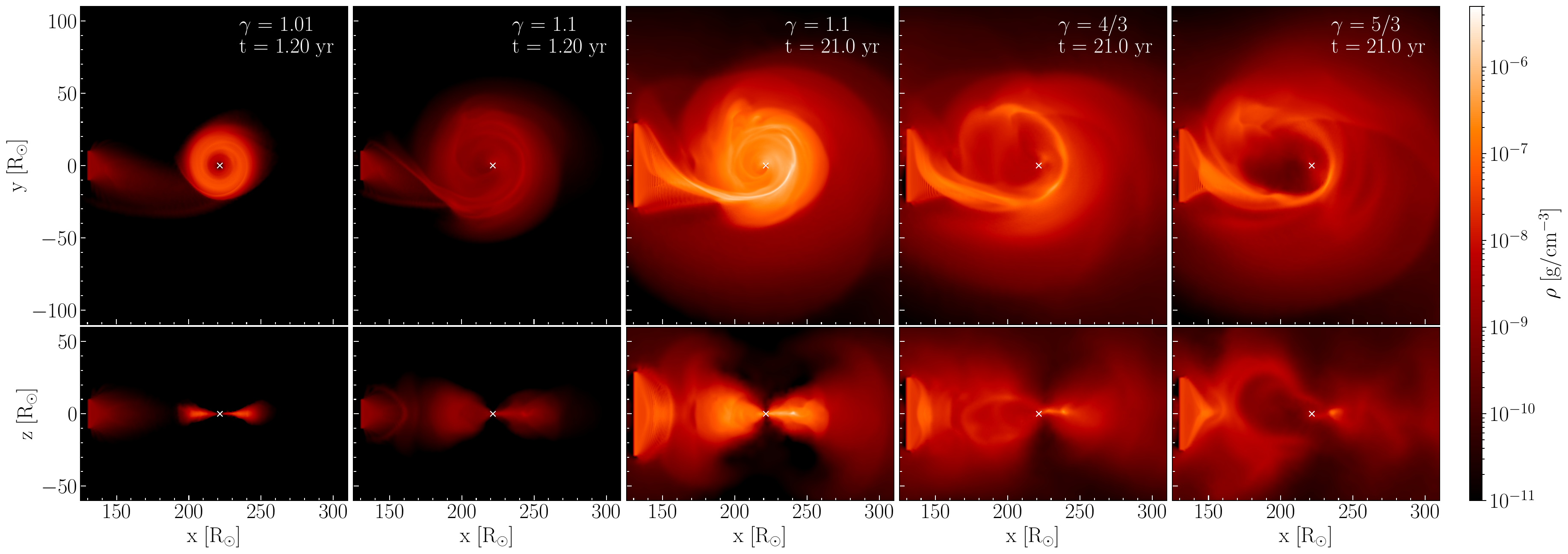}
    \caption{Density cross-sections in the orbital plane (top panels) and perpendicular planes of simulations with adiabatic indices $\gamma=1.01$ and $\gamma=1.1$ at 1.2~yr (first two columns) and $\gamma=1.1$, $4/3$, and $5/3$ at 21~yr (last three columns). The position of the accretor is marked with a cross.}
    \label{fig:gamma_4_3}
\end{figure*}

The first two columns of Figure~\ref{fig:gamma_4_3} show density slices from the $\gamma=1.01$ and $\gamma=1.1$ simulations at 1.2~yr. As expected, the greater compressibility of the $\gamma=1.01$ simulation leads to higher densities. Relative to the fiducial model, the resulting disk is thinner and nearly twice as massive. Its maximum density is approximately one order of magnitude higher, which substantially reduces the global Courant timestep and makes the simulation computationally expensive to evolve to 21~yr. 

The rightmost three columns of Figure~\ref{fig:gamma_4_3} show cross-sections of simulations performed with adiabatic indices of $\gamma=1.1$, 4/3 and 5/3 simulations at 21~yr. For the two highest values of $\gamma$, the injected material is deflected around the companion, producing a more extended, diffuse, and non-axisymmetric structure. In these cases, the stronger pressure response to compression prevents the formation of a disk. The corresponding edge-on views further demonstrate that the gas does not settle into a disk-like configuration. These results are consistent with those of Paper~I and \citet{Theuns1993,makita2000two, Huarte2013, macleod2015asymmetric, Murguia-Berthier2017AccretionFormation}, which showed that disk formation is highly sensitive to the gas adiabatic index. 

The fiducial value $\gamma=1.1$ was chosen because of its ability to produce a disk while maintaining inner-disk temperatures of the order of $\sim$1\e{7}~K, similar to those observed in NS X-ray binaries \citep{Mitsuda1994,Tauris2006}. For comparison, inner disk temperatures of $\sim 2\times10^{7}$~K and $\sim 9\times10^{6}$~K have been inferred for Cyg X-2 and Her X-1, respectively \citep{DiSalvo2002,Schandl1994,Revnivtsev2025}.

\section{Conclusion}
\label{sec:conclusion}

The formation of accretion disks via unstable \acrfull{RLOF} and their fate remain poorly understood in \acrshort{CE} simulations \citep{Macleod2018, MacLeod2020, Lau2022}. In this paper, we have studied the runaway \acrshort{RLOF} mass transfer phase in detail, including the formation of an accretion disk and the properties of the associated outflows. We have performed 3D hydrodynamic simulations using the SPH code \phant. The binary parameters are those presented in Paper~I, a 7~\msun\ \acrlong{RG} undergoing unstable mass transfer onto a 1.41~\msun\ neutron star. We model the final 21~yr of the \acrshort{RLOF} phase, just before a putative CE inspiral, where the mass transfer rate is predicted by a 1D \mesa\ model. During this period the mass transfer rate increases from 1.3~\e{-4}~\mdot\ to 1.0\e{-1}~\mdot.
 
The main results of our work are as follows:

\begin{enumerate}
    \item By the end of the simulation, an accretion disk with a mass of $\sim$5\e{-3}~\msun, a radius of $\sim$40~\rsun, and an outer disk scale height of $\sim 5$~\rsun\ has formed. The disk exhibits a sub-Keplerian structure, with a temperature profile of $T \propto r^{-1.3}$. By the end of the simulation at 21~yr, the mass accretion rate onto the companion reaches $\sim$5\e{-3}~\mdot. This accretion rate exceeds the \acrlong{NS} Eddington limit, and is expected to drive jets and outflows. Since these processes are not modelled in the simulation, the reported accretion rate should be treated as an upper limit.
        
    \item Using the thin-disk prescription of \citet{Shakura1973}, we estimate the effective viscosity parameter of the accretion disk to lie in the range $\alpha_{\eff}\simeq 0.03 - 0.06$. The upper limit is measured at the disk circularization radius, $r\simeq 25$~\rsun, beyond which the injection stream intersects the disk, rendering the thin-disk analysis invalid. To interpret the measured values of $\alpha_{\eff}$, we also compute the Reynolds stress parameter, $\alpha_{\rm R}$, which quantifies angular momentum transport driven by turbulent velocity fluctuations. We find $\alpha_{\rm R}\simeq 0.04 -0.11$ over the same radial range. The agreement between these two independent estimates and the absence of magnetic fields in the simulation indicate that the inferred accretion rate is consistent with angular momentum transport driven by hydrodynamical turbulence.

    \item The gas flowing out of the binary through $L_{2}$ forms a spiral pattern, carrying an average specific angular momentum $h_{\loss} = 0.87~h_{\Ltwo}$ (equivalently, $\gamma_{\loss} = h_{\loss} / h_{\orb} = 10.2$), where $h_{\Ltwo}$ is the specific angular momentum of a particle at $L_{2}$, and $h_{\orb}$ is the initial specific angular momentum of the binary. Both quantities remain approximately constant during the first $\sim5$~yr, and the total mass loss through outflows via $L_2$ corresponds to 14\% of the injected mass. 
    Since $h_{\loss}$ lies slightly below $h_{\Ltwo}$, mass and angular momentum are extracted efficiently from the binary, and indicate that isotropic re-emission is a poor approximation when mass transfer rates are high. Using the last simulated values of mass transfer rate, mass accretion rate and $\gamma_{\loss}$, we estimate that the binary would enter a \acrshort{CE} phase approximately 6~yr after the end of our simulation. If included in the simulations, jets could reduce the accretion rate onto the neutron star but would constitute only a small fraction of total mass loss, which is dominated by $L_{\rm 2}$ outflows. This is a conservative upper estimate, since in reality the mass transfer rate is growing faster, and the binary would enter the CE phase sooner.
        
    \item We verified the robustness of our disk simulations by comparing SPH and grid-based numerical methods. Using nearly identical initial conditions, we find overall agreement in the disk properties, with differences ranging from less than 1\% to 10\%. The only notable exception is the mass accretion rate, which differs by 45\%. In this case, the discrepancy may be attributed to the analysis method in addition to the numerical methods. Unlike the fully converged grid-based calculations of Paper I, the SPH calculations presented here are not yet fully converged.
    
    \item The SPH setup was improved compared to Paper~I, by implementing an optically thick mass-transfer prescription with a sonic injection velocity. These modifications produce only minor differences, with the largest being a $\sim 20$\% reduction in the disk scale height.
    
    \item We find that evolving the 3D simulations for different physical times (10, 21 and 45~yr) prior to the end of the mass transfer phase does affect the total mass injected to the simulation, while the disk parameters remain robustly consistent across simulations. The disk aspect ratio and radial $\alpha_{\rm eff}$ and $\alpha_{\rm R}$ profiles differ by less than 10\%. The disk mass and accretion rate vary up to a factor of two, due to the 10~yr simulation having less cumulative mass injected. Although modelling longer simulations would be beneficial for reaching a steady state, our results of disk properties and $\alpha-$parameters do not depend strongly on our choice of starting time. 
    
    \item We find that an adiabatic index of $\gamma \leq 1.2$ is required for an accretion disk to form in 3D simulations without explicit cooling, as also reported in Paper~I and in the literature \citep[e.g.,][]{makita2000two}.
\end{enumerate}

We have provided hydrodynamic simulations of the formation and evolution of an accretion disk during the binary's high-mass transfer phase. However, several improvements need to be implemented in future work: a cooling prescription, radiative transfer, magnetic fields, and super-Eddington mass loss mechanisms. Additionally, the donor star must be resolved simultaneously within the simulation domain. These improvements are essential to constrain our values of $\gamma_{\rm loss}$.

\paragraph{Acknowledgments}
This work was supported by resources awarded under Astronomy Australia Ltd’s ASTAC merit allocation scheme on the OzSTAR national facility at Swinburne University of Technology. The OzSTAR program receives funding in part from the Astronomy National Collaborative Research Infrastructure Strategy (NCRIS) allocation provided by the Australian Government, and from the Victorian Higher Education State Investment Fund (VHESIF) provided by the Victorian Government. This research was undertaken with the assistance of resources from the National Computational Infrastructure (NCI Australia), an NCRIS enabled capability supported by the Australian Government. AJG acknowledges funding support from Macquarie University through the International Macquarie University Research Excellence Scholarship (‘iMQRES’, \nolinkurl{https://doi.org/10.82133/C42F-K220}), the Postgraduate Research Fund (`PGRF’), and the funding support from ASA via the Student Travel Assistant Scheme. MYML acknowledges funding by the European Union (ERC, ExCEED, project number 101096243). Views and opinions expressed are however those of the authors only and do not necessarily reflect those of the European Union or the European Research Council Executive Agency. Neither the European Union nor the granting authority can be held responsible for them. LS is research director at the F.R.S.-FNRS. RN acknowledges funding from the Australian Research Council via FT250100748 and from UKRI/EPSRC through a Stephen Hawking Fellowship (EP/T017287/1).

\paragraph{Data Availability Statement}
The data underlying this article will be shared on reasonable request to the corresponding author. Initial PHANTOM snapshots, .in files and .setup files are available at Zenodo: doi:\url{10.5281/zenodo.20710411} 

\printendnotes

\bibliography{example}

\appendix
\renewcommand{\thefigure}{A\arabic{figure}} 
\setcounter{figure}{0}
\setcounter{equation}{0} 
\renewcommand{\theequation}{A\arabic{equation}} 
\setcounter{table}{0} 
\renewcommand{\thetable}{A\arabic{table}} 

\section{Mass injection through \texorpdfstring{$L_{1}$}{L1}}
\label{app:appendix_radius}

In this section, we discuss the numerical implementation of particle injection through the ``nozzle'', used to simulate the mass transfer from the donor star at the $L_{1}$ point. We adopt the optically thick mass transfer regime of \citet{Kolb90} where the donor overfills its Roche lobe. In the simulation, the mass transfer stream is injected via a nozzle at the $L_{1}$ point with a circular cross-sectional area given by

 \begin{equation}
    S_{\Lone} = \frac{2 \pi F_{\Lone}(q) R_{\Lone}^3}{G M_{d}} (\Phi_{\rm ph} - \Phi_{\Lone}),
    \label{eq:kolb}
\end{equation}
where $R_{\Lone}$ is the Roche radius of the donor star, $\Phi_{\rm ph}$ and $\Phi_{\Lone}$ are the gravitational potentials at the donor's photosphere and the $L_1$ point, respectively, and $F_{\Lone}(q)$ is a dimensionless function of mass ratio that accounts for the geometry of the nozzle \citep{Ritter1988} that is given by
\begin{equation}
    F_{\Lone}(q) =1.23 + 0.5\log(q),
\end{equation}
with the mass ratio defined as $q= M_{a}/M_{d}$. Equation~\ref{eq:kolb} is numerically integrated within the mass transfer subroutine (binary module) in the 1D \mesa\ simulation. We modified the file \texttt{run\_binary\_extras.f90} to output the nozzle area $S_{\Lone}$ and to compute the equivalent circular radius, $(S_{L1}/\pi)^{1/2}$. 

\begin{figure}
    \centering
    \includegraphics[width=0.85\columnwidth]{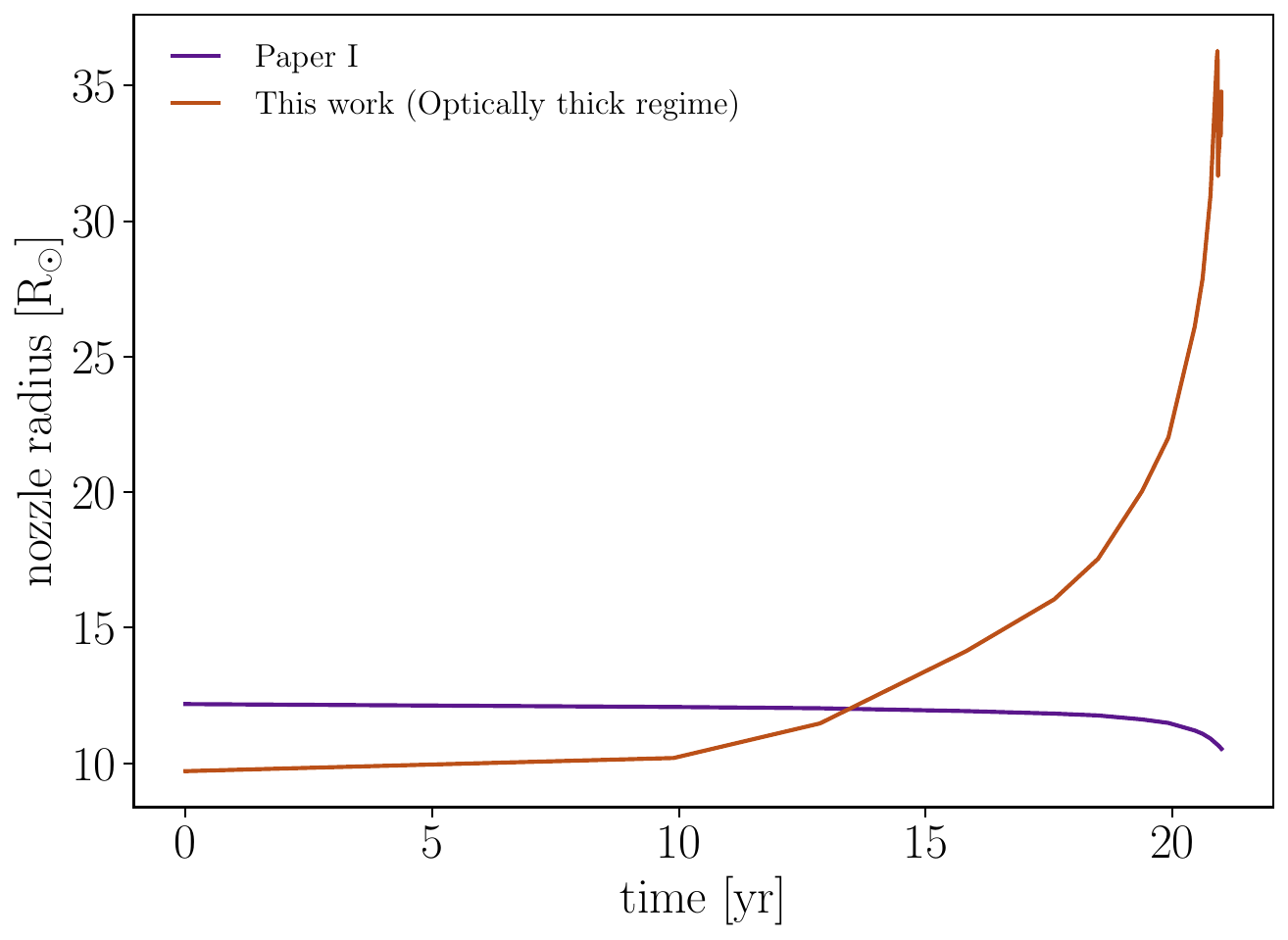}
    \caption{Nozzle radius evolution using the optically thin expression from \citep{Ritter1988} and the optically thick mass transfer expression of \citep{Kolb90} over the last 21~yr of the mass transfer phase of the \mesa\ simulation.}
    \label{fig:app_rad_nozz}
\end{figure}

Figure~\ref{fig:app_rad_nozz} shows the evolution of the nozzle radius over the 21~yr of the 3D simulation, comparing the nozzle radius prescription derived for optically thick mass transfer (red line) to the prescription for optically thin mass transfer used in Paper~I (purple line), when the mass transfer being optically thick throughout. Using the optically thick mass transfer regime, the nozzle radius grows significantly from 9.7~\rsun\ to 33.9~\rsun\ over 21~yr. As shown in Sect.~\ref{ssec:comp_nozz}, using a smaller or growing nozzle has little impact on the disk parameters.

The distribution of particles within the nozzle is obtained by assuming that the nozzle is a cylindrical region with its axis aligned along the line connecting the two stars (the $x$-axis). The injected particles are arranged in a cubic close-packed lattice along the cylinder, consisting of four evenly spaced layers, placed at the position of $L_{1}$. As explained in Section~\ref{sec:setup_3D}, the gas particles belonging to the nozzle are given fixed density ($\rho_{\Lone}$), velocity $v_{\Lone} = (c_{\rm s},0,0)$, and temperature (equal to the donor's photosphere temperature). This ensures the gas injection rate in the 3D simulation matches the mass transfer rate given by the \mesa\ simulation. 

Each time a layer leaves the nozzle region, a new layer is added at the back of the cylinder with updated density. The spacing between layers, and between particles within a layer, is determined by the required nozzle density at each time step. This density is derived from the mass transfer rate interpolated from the \mesa\ simulation.


\section{Merging SPH particles}
\label{app:merging}

\begin{figure*}[ht!]
    \centering
    \includegraphics[width=\columnwidth]{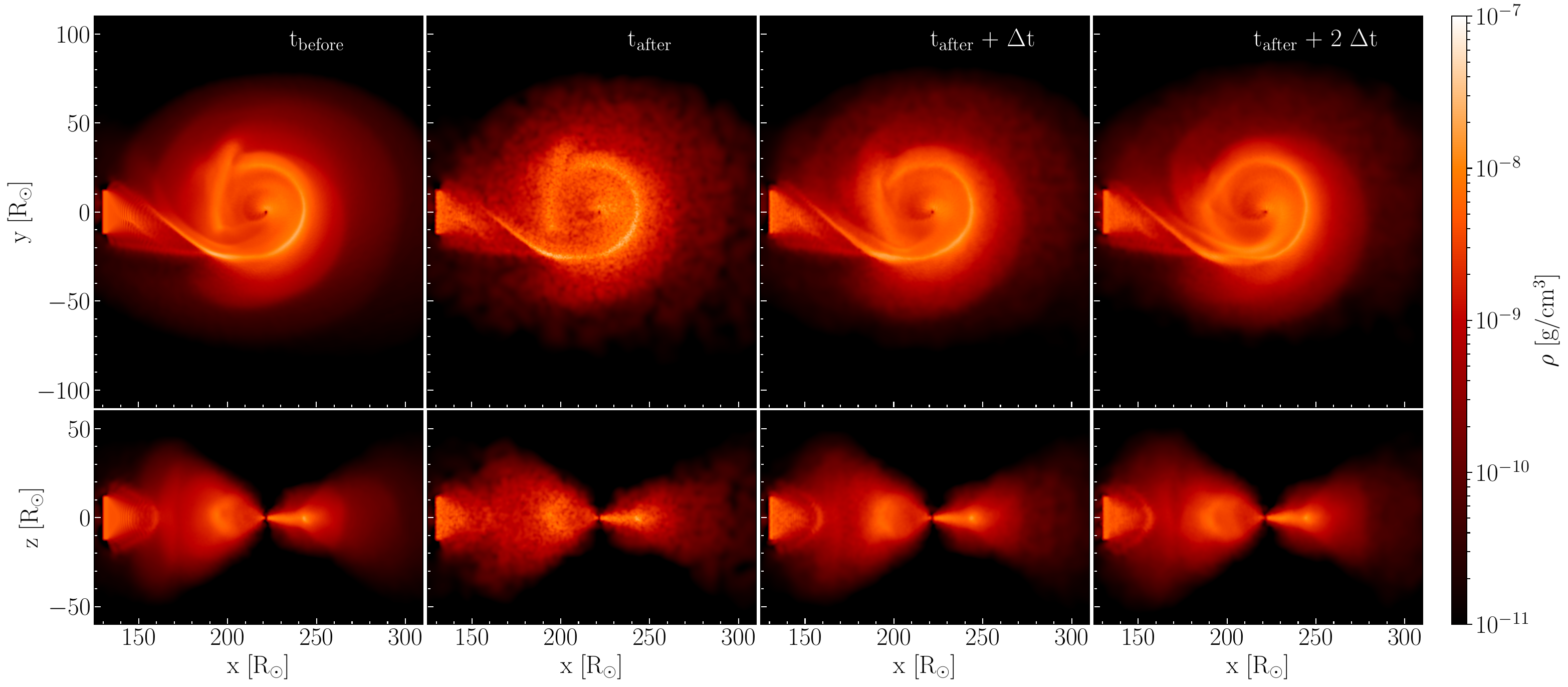}
    \caption{Density cross-sections in the orbital plane (top panels) and perpendicular plane (bottom panels) of simulation {\it sim-fiducial} at $t=13.3$~yr before merging SPH particles (first column) and after merging (second column). Third and fourth columns are 2 time steps after the merge, with $\Delta t = 5$\e{-3}~yr.}
    \label{fig:merging_density}
\end{figure*}    
The particle merging scheme developed by \cite{Nealon25} is used to reduce the number of particles when it exceeds some threshold. The code employs a modified k-d tree to identify and pair each particle with its nearest neighbour. For each pair, the \acrlong{COM} is calculated based on their positions. One particle is removed, while the remaining particle is given the position and velocity of the pair's \acrlong{COM}. The smoothing length of the new particle ($h_{\rm new}$) is scaled according to $h_{\rm new} = 2^{1/3} h$, where $h$ is the smoothing length of the remaining particle. Its mass is set to the sum of the particles' masses, ensuring mass conservation in the system.

We merged particles whenever the total number in the domain exceeded 2 million. This threshold allowed us to complete each simulation in Table~\ref{tab:simulations} in a few weeks.\footnote{Using 32 cores on a single Intel Xeon Gold processor.} In Figure~\ref{fig:merging_density} we visually inspect an SPH particle merge event by showing density cross-sections on the orbital (top) and perpendicular (bottom) planes for simulation {\it sim-fiducial}. We compare the disk's state at $t_{\rm before}~=~$13.3~yr (with 2\,089\,434 SPH particles), with the disk's state immediately following the merge at $t_{\rm after}$ (with 1\,044\,717 SPH particles), and two subsequent snapshots at intervals of $\Delta t~=~$5\e{-3}~yr. The particle merge leads to a visually coarser representation of the disk density, consistent with \citet{Nealon25}. Nevertheless, the global disk morphology, including the spiral structure, disk radial extent and vertical profile remain qualitatively unchanged. The absence of noisy particle distribution in snapshots following $t_{\rm after}$ suggests that the disk's large-scale evolution is not disrupted by the merging process. Although the visual inspection is qualitative, it complements the more rigorous convergence and mass transfer tests presented in the following sections.

\section{Resolution test}
\label{app:resolution_test}

\begin{figure*}[ht]
    \centering
    \includegraphics[width=0.9\columnwidth]{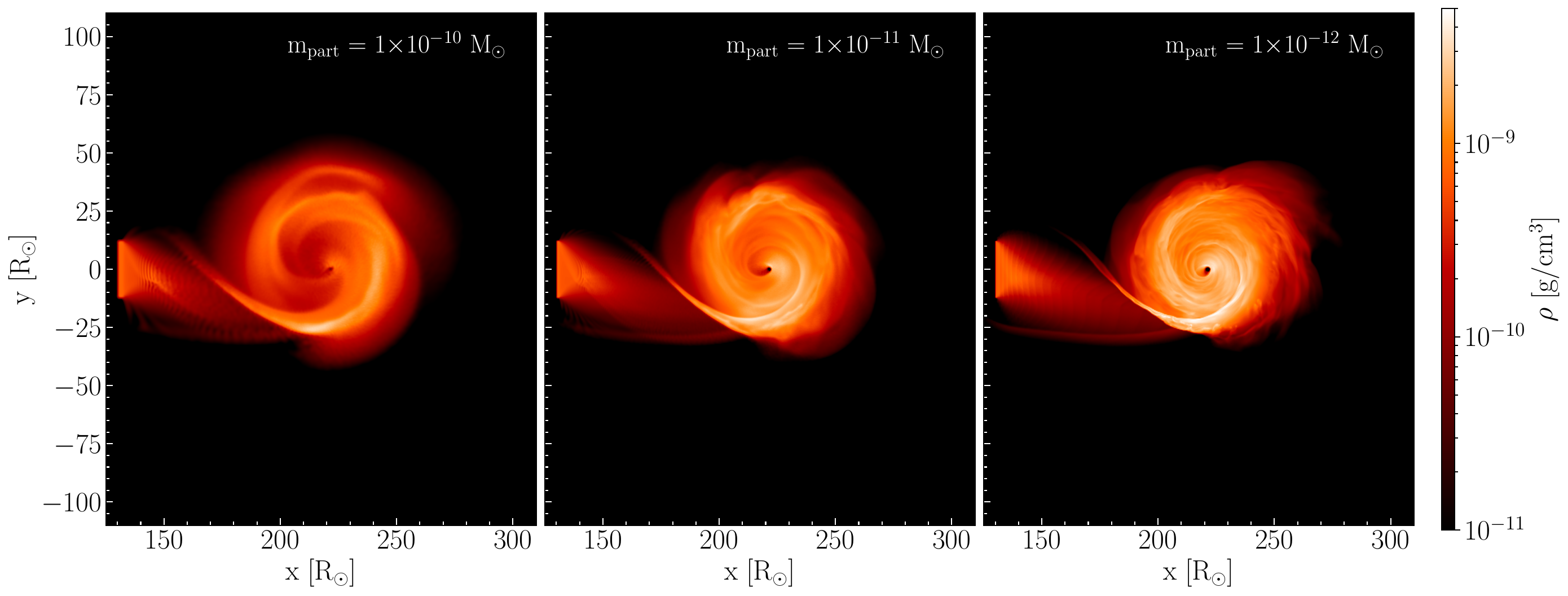}
    \caption{Density slices in the orbital plane at 0.5 yr with increasing resolutions (decreasing SPH particle mass). Left panel: lowest resolution with a particle mass of 1\e{-10}~\msun; middle panel: intermediate resolution with a particle mass of 1\e{-11}~\msun, and right panel: highest resolution with a particle mass of 1\e{-12}~\msun.}
    \label{fig:apx_renders}
\end{figure*}    

Because the fiducial simulation uses particle merging to keep simulation runtime within a few weeks, the resolution decreases over time, but the constant mass injection increases the total mass in the simulation. This makes it difficult to conduct a convergence test. We carried out a one year simulation similar to the fiducial simulation, using three particle masses: 1\e{-10}~\msun, 1\e{-11}~\msun, and 1\e{-12}~\msun, with no particle merging. All simulations have an injection velocity of $v_{\rm inj}\,=\,c_{\rm s}$ and a constant mass transfer rate of 1$\times 10^{-4}$~\mdot. The simulation with particle mass 1\e{-12}~\msun\ was evolved for only 0.5~yr, at which point it contained 50 million SPH particles and had slowed to the point where we were unable to evolve it further.

Figure~\ref{fig:apx_renders} shows the density slices on the orbital plane after 0.5 years for these three simulations. For all the simulated resolutions, an accretion disk forms. The injection layers at the nozzle are visible in the low resolution, 1\e{-10}~\msun\ particle mass simulation, while in the lower particle mass simulations we see a smoother injection stream. In the highest resolution simulation (1\e{-12}~\msun\ particle mass), the layered appearance of the injection stream is an artifact caused by the mass injection procedure resetting when the simulation is restarted. At all resolutions, high-density spiral arms develop at radii beyond $\sim30$~\rsun\ as a result of the interaction between the injection stream and the accretion disk. The highest resolution simulation resolves small scale structures within the disk that are smoothed out at the lowest resolution. A similar test with even lower resolution ($m_{\rm part}$ = 1\e{-9}~\msun) did not form a disk after 1 year, though it may form a disk if evolved further. 

\begin{figure}[ht]
    \centering
    \includegraphics[width=0.85\columnwidth]{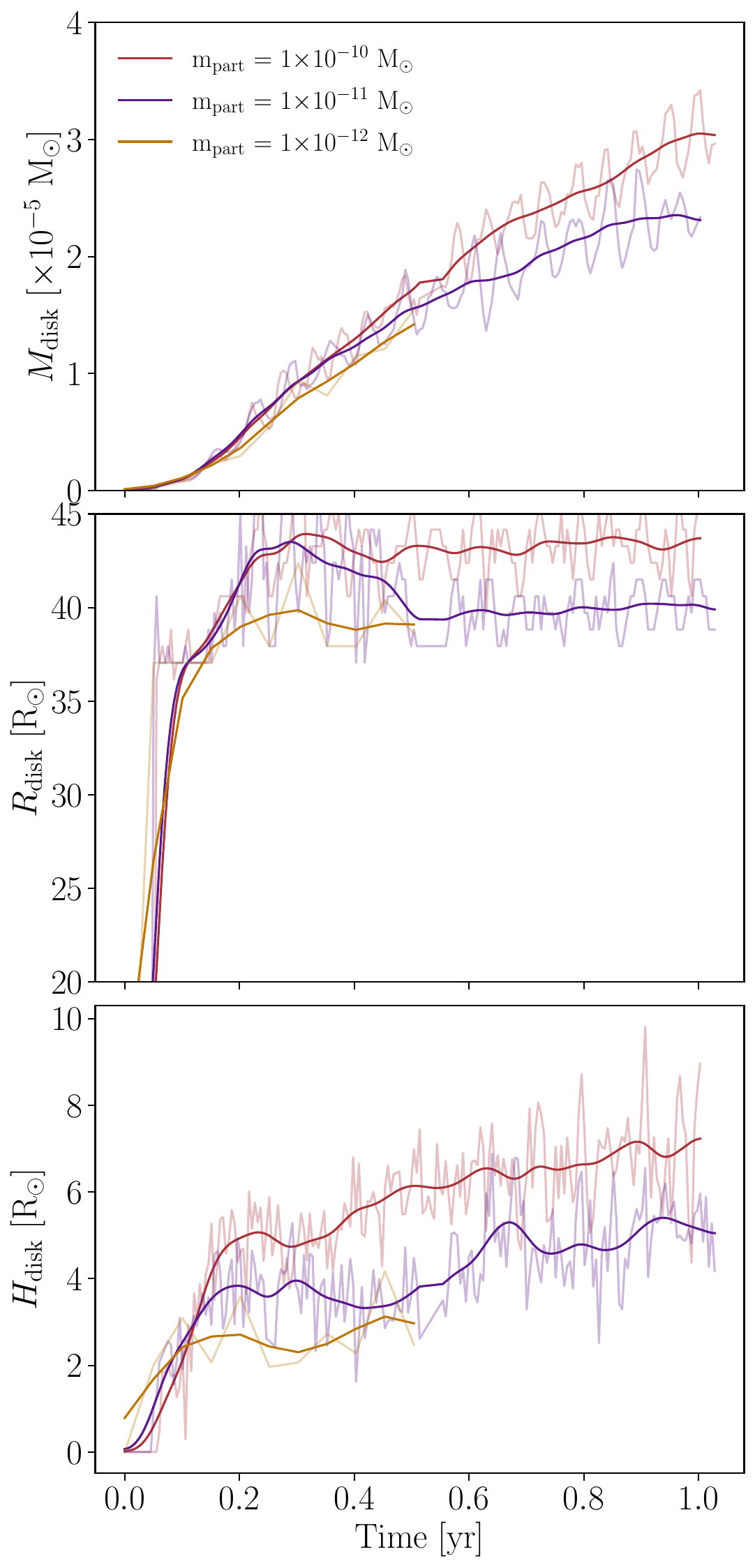}
    \caption{Evolution of selected disk properties for three different resolutions, corresponding to particle mass of 1\e{-10}, 1\e{-11} and 1\e{-12}~\msun\ (red, purple, and yellow curves, respectively). Solid lines show smoothed values. Top panel: disk mass evolution, middle panel: disk's radius evolution, and bottom panel: disk scale height evolution.}
    \label{fig:app_disk_evolution}
\end{figure}

The evolution of the disk parameters (mass, radius and scale height) for the three simulations is shown in Figure~\ref{fig:app_disk_evolution}. At 0.5~yr, for the two highest resolution simulations ($m_{\rm part} = 1\times10^{-11}$~\msun\ and $m_{\rm part} = 1\times10^{-12}$~\msun), the disk's radius agrees to within $\sim1.5$\%. The disk mass differs by $\sim10$\%, and the disk scale height by $\sim16$\%.


   
\begin{figure}[ht]
    \centering
    \includegraphics[width=0.85\columnwidth]{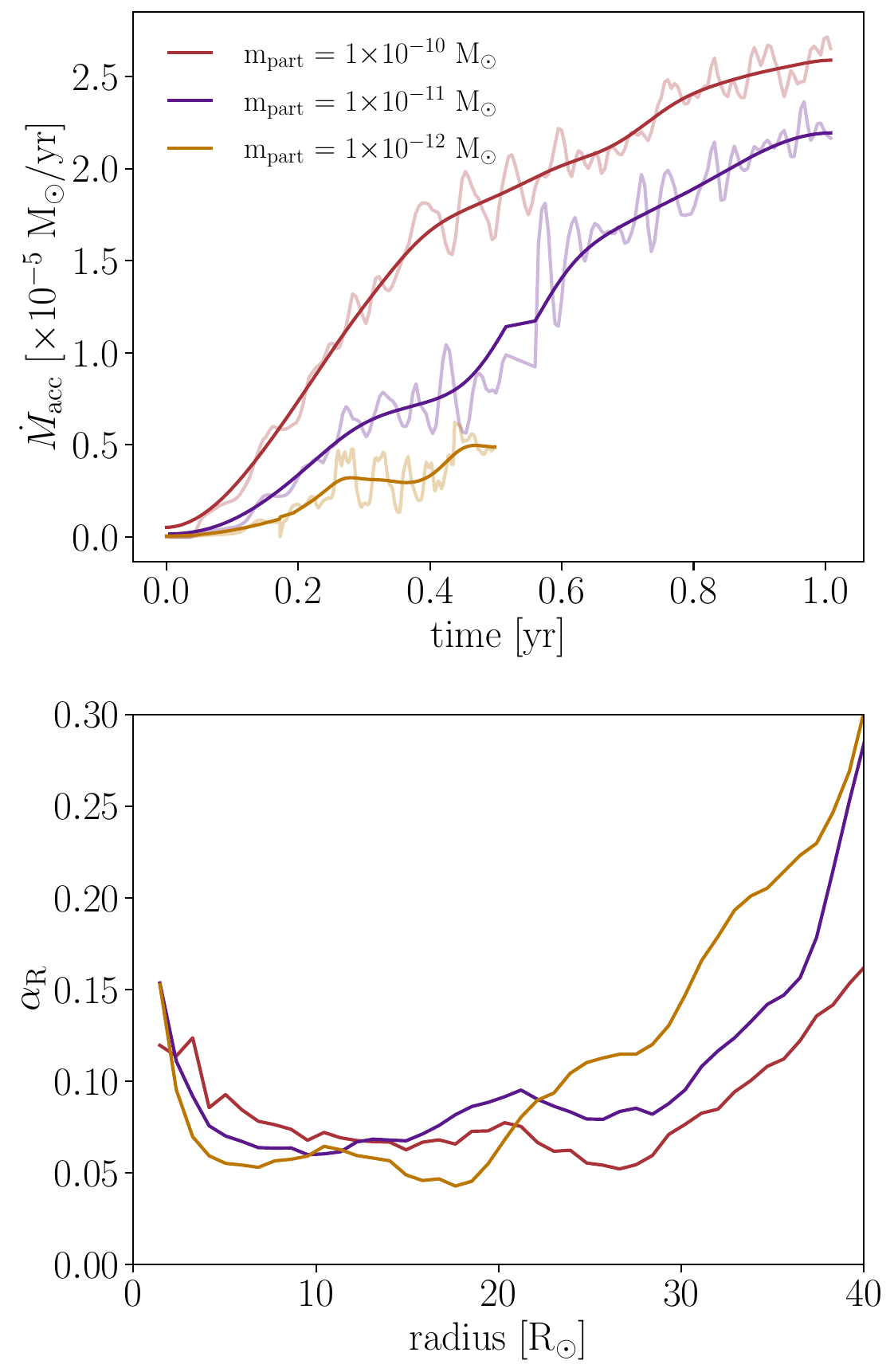}
    \caption{Top panel: smoothed mass accretion rate versus time for three different resolutions, particle mass of 1\e{-10}~\msun\ (red line), 1\e{-11}~\msun\ (purple line), and 1\e{-12}~\msun\ (yellow line). Bottom panel: short-term average $\alpha_{\rm R}$ parameter profile ($\delta t = 0.25$~yr) for three different resolutions.}
    \label{fig:app_mdot_alpaha_converge}
\end{figure}

The accretion rate onto the companion ($\dot{M}_{\rm acc}$, Figure~\ref{fig:app_mdot_alpaha_converge}, top panel) is found to decrease with increasing resolution. We find $\dot{M}_{\rm acc}\propto m_{\rm part} ^{0.27\pm 0.02}$, where $m_{\rm part}$ is the particle mass and $\dot{M}_{\rm acc}$ is measured at 0.5~yr. Because the SPH smoothing length scales as $h \propto m_{\rm part}^{1/3}$, this result implies that the accretion rate is approximately proportional to $h$. We note that the measurements are taken up to 0.5~yr, which is shorter than the simulation's viscous timescale ($t_\mathrm{visc} = 10 yr$), meaning the disk is not in steady state. 
In the bottom panel of Figure~\ref{fig:app_mdot_alpaha_converge}, we show the radial profile of short-term averaged ($\delta t =0.25$~yr) Reynolds stress parameter $\alpha_{\rm R}$, measured using Equation~\ref{eq:alpha}. As seen in Section~\ref{ssec:mdot_temp_vel_alpha}, $\alpha_{\rm R}$ increases steeply for radii larger than 25~\rsun, near the circularization radius. Despite the stochastic behaviour, the $\alpha_{\rm R}$ values across all three resolutions are similar, remaining between $\sim$0.01 and $\sim$0.15 for $r \lesssim 25$~\rsun, before increasing to $\sim 0.3$ at larger radii.

\section{Measurement of \texorpdfstring{$\alpha$ and $h_{\mathrm{loss}}$}{alpha and h\_loss}}
\label{app:alpha_and_am_loss}

The determination of the $\alpha$ viscosity parameters and specific angular momentum, $h_{\loss}$, carried away by the outflow, (Sections~\ref{ssec:mdot_temp_vel_alpha} and \ref{ssec:turbulence_angular_momentum}), relies on spatial averaging of SPH quantities within concentric cylindrical shells. The adopted shell widths and control surface locations could potentially affect the inferred values. To quantify this effect, we varied the number of sampling shells and their radial extents.

\begin{figure}[ht]
\centering
\includegraphics[width=0.8\columnwidth]{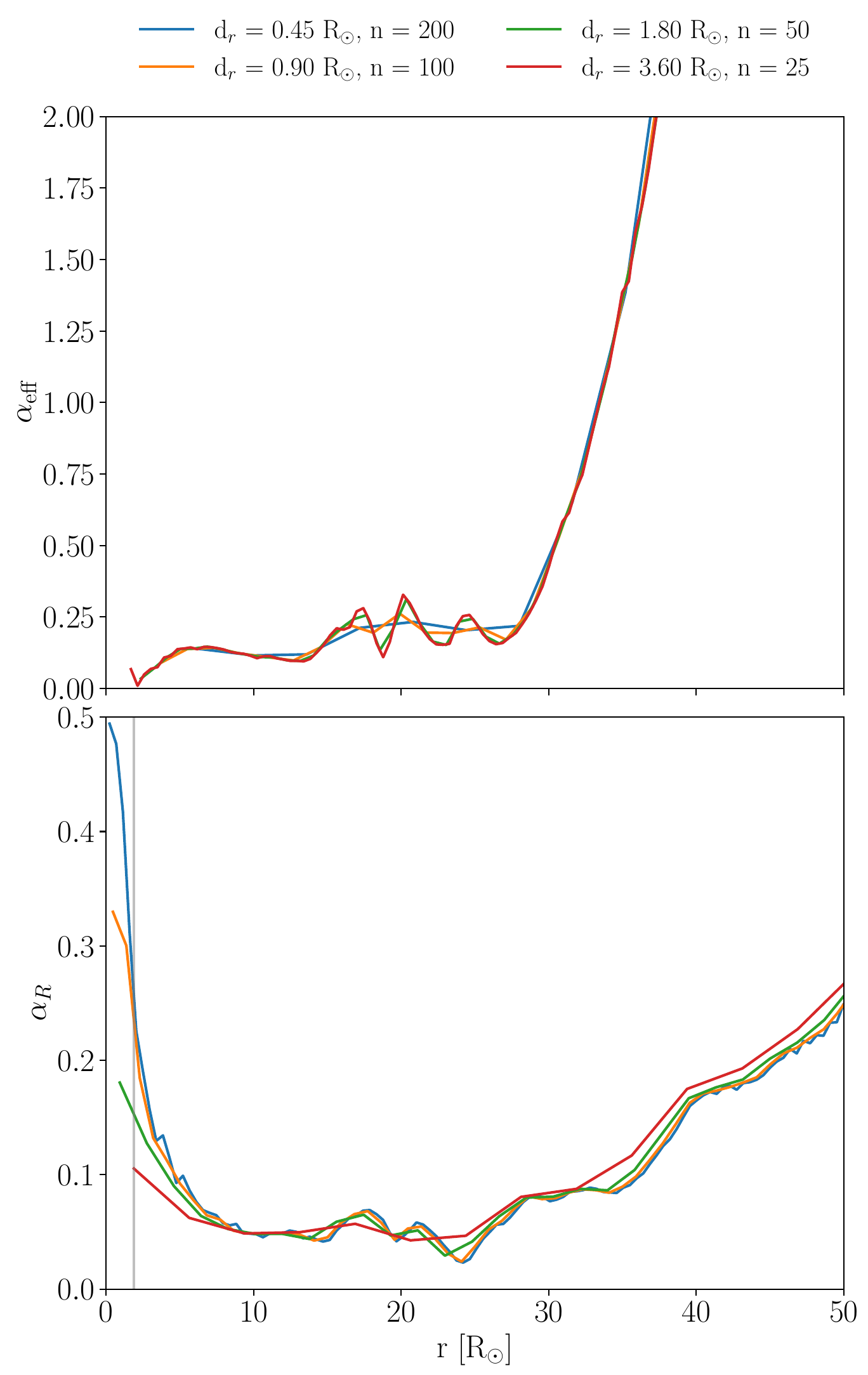}
\caption{The $\alpha$ parameters at the end of the fiducial simulation (21 yr). Top panel, $\alpha_{\eff}$ calculated with Eq.~\ref{eq:alpha_ss}; Bottom panel, $\alpha_{\rm R}$ measured with Eq.~\ref{eq:alpha} - the grey line shows the accretion radius. The different coloured lines correspond to calculating the average $\alpha$ with a different number of radial bins.}
\label{fig:app_alpha}
\end{figure}

The $\alpha$ parameters were initially calculated using 100 concentric cylindrical shells each with a radial width of $d_r = 0.9$~\rsun. In Figure~\ref{fig:app_alpha}, we compare the radial profiles of $\alpha_{\eff}$ (left panel) and $\alpha_{\rm R}$ (right panel) at 21~yr for 25, 50, 100 and 200 shells, which have decreasing shell widths of 3.60, 1.80, 0.90 and 0.45~\rsun, respectively. The $\alpha_{\eff}$ and $\alpha_{\rm R}$ profiles show small variations across different bin sizes, confirming that our spatial averaging is well converged.

We also tested the method used to determine the specific angular momentum, $h_{\loss}$, of the gas leaving through $L_{2}$ (Section~\ref{ssec:turbulence_angular_momentum}). We computed the mass and angular momentum fluxes using two control surfaces: (i) a spherical shell of radius $r_{\rm sphere}$ = 1500~\rsun\ centered on the \acrshort{COM}, with thickness $d_r~=~1$~\rsun, and (ii) a smaller cylindrical sector located at $r_{\rm cylinder}$ = 1.1$r_{\Ltwo}$ with the same thickness, angular extent $\phi \in [-\pi/2, \pi/4]$, and vertical range of $z \in [0, 1.1r_{\Ltwo}]$. In both cases, only particles crossing the control surface with outward radial velocities were included in the flux calculation.

\begin{figure}[ht]
\centering
\includegraphics[width=0.8\columnwidth]{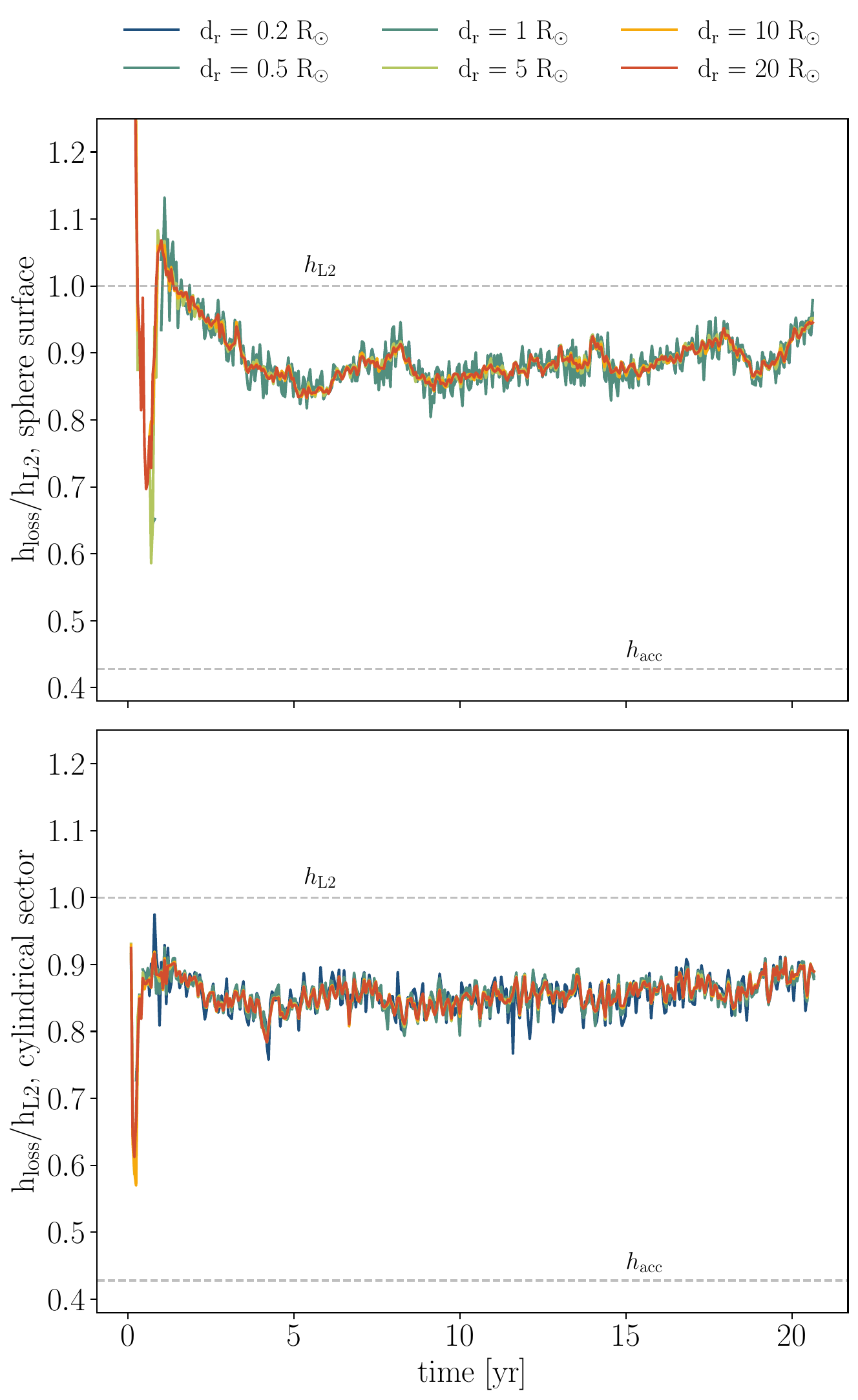}
\caption{Specific angular momentum loss, $h_{\loss}$, for outflowing material. Top panel: Impact of varying the shell width (d$_r$) for a fixed radial position at r$_{\rm shell}$= 1500~\rsun. Bottom panel: same as top panel but for the method using the cylindrical sector in the vicinity of the $L_2$ point.}
\label{fig:app_outflowing}
\end{figure}

To assess the sensitivity of our results to the thickness of the control surface, we show $h_{\loss}/ h_{\Ltwo}$ in Figure~\ref{fig:app_outflowing} for the spherical shell method keeping its radius constant with $r_{\rm sphere} = 1500$~\rsun, but varying the shell thickness. We repeated the same analysis for the cylindrical-sector method. The  results show excellent agreement across all shell thicknesses in both cases, confirming that the measured specific angular momentum is converged with respect to the control-surface thickness and validating our adoption of a 1~\rsun\ shell.

\end{document}